\documentclass[%
    10pt,
    reprint,
    preprintnumbers,
    nofootinbib,
    amsmath,
    amssymb,
    aps, 
    prd,
    showkeys,
]{revtex/revtex4-2}

\usepackage{revtex/aas_macros}
\usepackage{bm}
\usepackage{booktabs}
\usepackage{graphicx}
\usepackage{dcolumn}
\usepackage{enumitem}
\usepackage[shortcuts]{extdash}
\usepackage[colorlinks,citecolor=OliveGreen]{hyperref}
\usepackage[scale=.75]{miama}
\usepackage{needspace}
\usepackage{outlines}
\usepackage{pdfrender}
\usepackage{physics}
\usepackage[separate-uncertainty,free-standing-units,list-units=single]{siunitx}
\usepackage{subfigure}
\usepackage{tabularx}
\usepackage[normalem]{ulem}
\usepackage[dvipsnames]{xcolor}
\usepackage{xspace}
\usepackage[nameinlink,capitalise]{cleveref}

\crefname{bayes}{Bayes' theorem}{Bayes' theorem}
\crefname{occam}{Occam equation}{Occam equation}
\creflabelformat{bayes}{#2(Eq.~#1)#3}
\creflabelformat{occam}{#2(Eq.~#1)#3}

\makeatletter
\newcommand\footnoteref[1]{\protected@xdef\@thefnmark{\ref{#1}}\@footnotemark}
\makeatother

\makeatletter
\g@addto@macro\bfseries{\boldmath}
\makeatother

\DeclareFontFamily{OT1}{mathc}{}
\DeclareFontShape{OT1}{mathc}{m}{it}{<-> mathc10}{}
\DeclareMathAlphabet{\mathcal}{OT1}{mathc}{m}{it}
\DeclareFontEncoding{LGR}{}{}
\renewcommand{\Pi}{\mathord{\text{%
    {\pdfrender{TextRenderingMode=FillStroke,LineWidth=0.15pt}{\fontencoding{LGR}\fontfamily{fmm}\selectfont P\,}}%
}}}

\AtBeginDocument{\RenewCommandCopy\qty\SI}
\DeclareSIUnit\parsec{pc}
\DeclareSIQualifier\planck{Pl}
\DeclareSIUnit\planckmass{\m\planck}
\DeclareSIUnit\plancktime{\t\planck}
\DeclareSIUnit\plancklength{\ell\planck}

\DeclareMathOperator\Erfc{Erfc}
\DeclareRobustCommand{\Planck}{\textsc{Planck}\xspace}
\DeclareRobustCommand{\LiteBIRD}{\textsc{LiteBIRD}\xspace}
\DeclareRobustCommand{\Cosmoglobe}{\textsc{Cosmoglobe}\xspace}
\DeclareRobustCommand{\PantheonPlus}{\textsc{Pantheon}\ensuremath{^+}\xspace}
\DeclareRobustCommand{\NuFIT}{\textsc{NuFIT}\,6.0\xspace}
\DeclareRobustCommand{\Lollipop}{\texttt{LoLLiPoP}\xspace}
\DeclareRobustCommand{\Hillipop}{\texttt{HiLLiPoP}\xspace}
\DeclareRobustCommand{\CamSpec}{\texttt{CamSpec}\xspace}
\DeclareRobustCommand{\Plik}{\texttt{Plik}\xspace}
\DeclareRobustCommand{\SimAll}{\texttt{SimAll}\xspace}
\DeclareRobustCommand{\Commander}{\texttt{Commander}\xspace}
\DeclareRobustCommand{\LCDM}{\ensuremath{\Lambda\mathrm{CDM}}\xspace}
\DeclareRobustCommand{\nuLCDM}{\ensuremath{\Lambda\mathrm{CDM}+\sum{m_\nu}}\xspace}
\DeclareRobustCommand{\KLCDM}{\ensuremath{\Lambda\mathrm{CDM}+\Omega_K}\xspace}
\DeclareRobustCommand{\wCDM}{\ensuremath{w\mathrm{CDM}}\xspace}
\DeclareRobustCommand{\wwaCDM}{\ensuremath{w_0w_a\mathrm{CDM}}\xspace}
\DeclareRobustCommand{\Omegam}{\ensuremath{\Omega_\mathrm{m}}\xspace}
\DeclareRobustCommand{\OmegaK}{\ensuremath{\Omega_K}\xspace}

\newcommand{\av}[2][]{\left\langle #2\right\rangle_{#1}}
\newcommand{\avP}[1]{\av[\mathcal{P}]{#1}}
\newcommand{\avp}[1]{\av[\Pi]{#1}}

\begin{document}


\title{Consistency of standard cosmologies\texorpdfstring{\\}{ }using Bayesian model comparison and tension quantification}

\author{Lukas Tobias Hergt}
\email{Contact author: lukas.hergt@ijclab.in2p3.fr}
\author{Sophie Henrot-Versillé}
\author{Matthieu Tristram}
\affiliation{Universit\'e Paris-Saclay, CNRS/IN2P3, IJCLab, 91405 Orsay, France}
\author{Douglas Scott}
\affiliation{Department of Physics and Astronomy, University of British Columbia, BC\;V6T\,1Z1, Canada}

\date{\today}

\begin{abstract}
    We present a unified Bayesian assessment of model comparison and data-set consistency for \LCDM (cold dark matter plus a cosmological constant) and minimal extensions (neutrino mass, spatial curvature, constant or evolving dark energy) using cosmic microwave background~(CMB), baryon acoustic oscillation~(BAO), and type~Ia supernova~(SN) data.
    The major results are summarized in the first three figures.
    We quantify model preference with Bayesian evidence and assess consistency with complementary evidence- and likelihood-based diagnostics applied uniformly across data-set combinations.
    For the models considered, updated \Planck processing systematically improves internal CMB consistency (low-$\ell$ versus high-$\ell$, and primary CMB versus CMB lensing).
    The preference for a closed geometry and an associated ``curvature tension'' with BAO and/or CMB lensing are largely confined to earlier \Planck likelihood implementations and weaken substantially when using updated CMB processing and more recent BAO measurements.
    Apparent evidence for evolving dark energy in CMB+BAO+SN combinations depends sensitively on the specific pairing of CMB and SN likelihoods: plausible alternatives shift inferred tensions by more than $\SI{1}{\sigma}$ and can completely reverse the preferred model.
    Allowing a free neutrino mass tends to absorb residual shifts without introducing new inconsistencies, and we do not find robust evidence for a standalone $\tau$-driven discrepancy once the full likelihood context is accounted for.
    We conclude that claims of a required update of our standard cosmological model from \LCDM to \wwaCDM are premature.
\end{abstract}

\maketitle

\section{Introduction}
\label{sec:introduction}

``Concordance'' cosmology emerged as a watchword, because independent constraints from the cosmic microwave background~(CMB), large-scale structure~(LSS) and other data sets converged on a particular model through the 1990s \cite{Efstathiou1990_SM,KraussTurner1995,OstrikerSteinhardt1995,OstrikerSteinhardtNature,Bahcall1995_dark_matter}. Type~Ia supernovae~(SN) joined these precision data sets and LSS constraints became more exacting through measurements of the baryon acoustic oscillations~(BAO).  CMB, BAO, and SN data sets, each leaving parameter degeneracies when considered on their own, intersect at roughly the same point in parameter space. This convergence favors a model that is close to spatially flat when allowing for curvature, and close to a pure cosmological constant~$\Lambda$ when varying the equation-of-state parameter~$w$ of the dark energy.
The same set of observations also converged on a small number of qualitative statements that have since become part of the standard cosmological narrative, in particular: structure formation requires a dominant, non-baryonic, dynamically cold, and nearly collisionless dark matter component; and the late-time expansion history requires a negative-pressure component consistent with $\Lambda$.

This basic picture has proven remarkably stable, and it underpins what has emerged as ``the standard model of cosmology'', usually referred to as \LCDM, shorthand for a Universe containing cold dark matter~(CDM) and dark energy in the form of a cosmological constant~$\Lambda$; see Ref.~\cite{Scott2018_skeptics_guide} for a concise discussion of what is meant by ``the standard model of cosmology'', why it is compelling, and why one should nonetheless remain appropriately skeptical.
With a small set of assumptions, \LCDM provides a successful joint description of a wide range of observables with just over a handful of parameters.
At the same time, as the statistical errors have contracted and systematic effects have become comparatively more important, the question of whether different data sets (and different implementations of ostensibly the same data set) agree with one another has become part of the inference problem: if they do not agree, then the interpretation of any combined result becomes ambiguous.

A substantial fraction of recent cosmology literature can be read as an attempt to interpret mild but persistent discrepancies within, or between, cosmological probes (for recent reviews, see Refs.~\cite{CosmoVerseNetwork2025,Ong2025_tension_grid}).
The most widely discussed example remains the Hubble (or $H_0$) tension between early-Universe inferences and late-time distance-ladder determinations calibrated with Cepheid variables~\cite{Riess2020_hubble_tension}.
However, there are other (typically lower-significance) discrepancies, including internal \Planck consistency questions (e.g., low\=/$\ell$ versus high\=/$\ell$ splits, or CMB power spectra versus CMB lensing reconstruction), the \Planck~2018 data-set curvature preference and its interplay with BAO data (the so-called curvature tension)~\cite{DiValentino2020_curvature,Handley2021_curvature,DiValentino2021_curvature}, and more recently renewed discussions of compatibility between CMB and BAO data in light of BAO updates from the Dark Energy Spectroscopic Instrument~(DESI) and their implications for dark-energy parameterizations~\cite{Choudhury2024_12parameters,Gu2025_desi_tension,Ye2025_desi_tension,Ferreira2025_cmb_bao_tension_inflation,Lee2025_desi,Ong2025_desi}.

A recurring theme in all of these examples is that the qualitative conclusions are often sensitive both to the choice of data products and their combinations, e.g., different conclusions about the lensing anomaly for different \Planck releases~\cite{Planck2018_parameters,CamSpec4_Rosenberg2022,Hillipop_Tristram2023}, or the choice of calibrator for the distance ladder, such as cepheids~\cite{Leavitt1912_cepheids,Hertzsprung1913_cepheids,Riess2020}, carbon rich stars~\cite{Richer1981_carbon_stars,Battinelli2005_carbon_stars,Ripoche2020_carbon_stars}, or the tip of the red giant branch~\cite{Lee1993_trgb,Freedman2020}.
For a fixed data-set choice, one expects different reasonable diagnostics of consistency to give broadly consistent qualitative inference results, differing mainly in their quantitative calibration.
However, the various statistics commonly grouped under the label ``tensions'' do not always answer the same question: fit-based measures, posterior-shift measures, and evidence-based criteria weight prior volume and partially shared constrained subspaces differently, making a comparison of claimed tensions based on different metrics difficult.
In this work we have therefore set out to adopt a consistent Bayesian framework across a wide range of data-set combinations and report complementary quantities, so that changes in conclusions can be traced to either the data inputs or to the specific hypothesis being tested.

We present a unified assessment of Bayesian model comparisons and data-set tensions for \LCDM and four minimal extensions.
We focus on the interplay of three types of observable: CMB and CMB lensing from \Planck (contrasting different data releases and high\=/$\ell$ likelihood implementations); BAOs from the Sloan Digital Sky Survey~(SDSS) for comparison with past findings, and from the recent data release of DESI; and SNe from the Pantheon and Dark Energy Survey~(DES) collaborations.
Our aim is not so much to advocate a particular extension, but to isolate which specific likelihood choices and data-set combinations drive apparent tensions or model preferences, and which of these features persist under updates of the underlying likelihoods. Nevertheless, we end up concluding that the combination of statistical tests suggests that there is no strong reason to consider that \LCDM is in need of revision.

Our paper is structured as follows. In \cref{sec:methods} we review the relevant Bayesian statistics and present an overview of the theoretical models and all data sets used in our analysis. The parameters going into the theoretical models and a list of all likelihoods are summarized in \cref{tab:priors,tab:likelihood_overview}, respectively. Our main results are presented in \cref{fig:tension_R,fig:tension,fig:model_comparison} and discussed in \cref{sec:results}. We conclude in \cref{sec:conclusion}.

\section{Methods}
\label{sec:methods}

Although Bayesian inference is well known to most cosmologists, in this paper we are going to discuss some details that may be less familiar, and hence we will start with a general introduction to the topic.

\subsection{The three pillars of Bayesian inference}

The late David MacKay identified two pillars (or levels) of Bayesian inference in his book ``Information Theory, Inference, and Learning Algorithms''~\cite{MacKayBookCh28}: (i)~parameter estimation; and (ii)~model comparison. With the sub-percent level precision of current cosmology experiments a third pillar has gained importance: (iii)~data-set consistency and tension quantification.

All three pillars are built on Bayes' theorem,
{\crefalias{equation}{bayes}%
\begin{alignat}{3}
    \Pr(\theta|D,M)       &\times \Pr(D|M)         & &=\,\, & \Pr(D|\theta,M)       &\times \Pr(\theta|M) , \nonumber \\
    \mathrm{Posterior}    &\times \mathrm{Evidence}& &=\,\, & \mathrm{Likelihood}   &\times \mathrm{Prior} , \nonumber \\
    \label{eq:bayes_theorem}
    \mathcal{P}(\theta) &\times \mathcal{Z}    & &=\,\, & \mathcal{L}(\theta) &\times \Pi(\theta) . 
\end{alignat}
}%
Assuming some model~$M$ (e.g., \LCDM or a minimal extension), and given the data~$D$, it relates the output quantities, the posterior distribution~$\mathcal{P}$ of a parameter set~$\theta$ and the Bayesian evidence~$\mathcal{Z}=\avp{\mathcal{L}}$, to the input quantities, the likelihood function~$\mathcal{L}$, and the prior distribution~$\Pi$.
Here we use $\avp{\cdot}$ and $\avP{\cdot}$ to denote prior and posterior averages, respectively.

We can re-arrange \cref{eq:bayes_theorem} to obtain the \emph{generalized Occam equation}:
{\crefalias{equation}{occam}%
\begin{alignat}{3}
    \label{eq:occam_equation_general}
    \text{(log-)evidence} &=\; & \text{parameter fit}\,   &- \,\text{Occam penalty} ,\nonumber \\
    \ln\mathcal{Z}  &= & \ln\mathcal{L}(\theta) &- \mathcal{I}(\theta),
\end{alignat}
}%
which decomposes the overall model quality in the form of the log-evidence~$\ln\mathcal{Z}$ into a goodness-of-fit term (the log-likelihood~$\ln\mathcal{L}(\theta)$) and the pointwise Shannon information $\mathcal{I}(\theta)=\ln[\mathcal{P}(\theta)/\Pi(\theta)]$ acting as an Occam penalty.

There are different options for how to handle the free parameter set~$\theta$ in the \cref{eq:occam_equation_general}. For example, one could choose the maximum likelihood estimate (i.e., the best-fit point)~$\hat\theta_\mathrm{ML}$, the posterior mode~$\hat\theta_\mathrm{MP}$, or the posterior mean~$\bar\theta$: 
\begin{align}
    \hat\theta_\mathrm{ML} &= \underset{\theta}{\max}\,\mathcal{L}(\theta);\\
    \hat\theta_\mathrm{MP} &= \underset{\theta}{\max}\,\mathcal{P}(\theta);\\
    \bar\theta &= \avP{\theta}.
\end{align}
The most Bayesian approach is to take the posterior mean of the entire equation,\footnote{Note that the left hand side of the \cref{eq:occam_equation_general} does not depend on the parameter set~$\theta$, such that $\avP{\ln\mathcal{Z}}=\ln\mathcal{Z}$.} arriving at the \emph{special Occam equation} that relates the Bayesian log-evidence to the Kullback--Leibler divergence {$\mathcal{D}=\avP{\mathcal{I}}$}, as already pointed out in Ref.~\cite{Hergt2021}:
{\crefalias{equation}{occam}%
\begin{equation}
    \label{eq:occam_equation_special}
    \ln\mathcal{Z}  = \avP{\ln\mathcal{L}} - \mathcal{D}.
\end{equation}
}%
The Kullback--Leibler divergence is the average contraction from prior to posterior, and also referred to as ``information gain'' or ``relative entropy''.

The other term in the \cref{eq:occam_equation_special}, the posterior average of the log-likelihood~$\avP{\ln\mathcal{L}}$, gives a measure of the fit of the model to the data. Similarly, the posterior \emph{variance} of the log-likelihood quantifies the model's flexibility, and is also referred to as the Bayesian model ``dimensionality'',
\begin{equation}
    \label{eq:model_dimensionality}
    \frac{\mathcal{d}}{2} = \avP{(\ln\mathcal{L})^2} - \avP{\ln\mathcal{L}}^2.
\end{equation}
This quantifies the number of degrees of freedom \emph{constrained} by the data, so any parameter that does not affect the likelihood will not augment the model dimensionality (for details, see Ref.~\cite{Handley2019_dimensionality}).

\subsubsection{Pillar 1: Parameter estimation}
\label{sec:methods_parameter_estimation}

The first level of Bayesian inference is dedicated towards finding the posterior distribution~$\mathcal{P}$ of a parameter set~$\theta$. At this stage the evidence is typically viewed as a mere normalization constant, and therefore neglected, such that only the proportionality
\begin{equation}
    \mathcal{P}(\theta) \propto \mathcal{L}(\theta) \times \Pi(\theta)
\end{equation}
is of interest.

When reported, posterior distributions are often summarized by their posterior averages and credible regions of the marginalized parameters. Plotting the marginalized one-dimensional distributions, or the two-dimensional contours, allows for a slightly better representation of the underlying distribution, but they can nonetheless suffer from projection effects (see e.g., figure~7 in Ref.~\cite{Handley2019_tension}).

\subsubsection{Pillar 2: Model comparison}
\label{sec:methods_model_comparison}

At the second level of Bayesian inference, we are interested in comparing different models, say $M_1$ and $M_2$, with one another given the same data~$D$:
\begin{equation}
    \label{eq:model_comparison}
    \frac{\Pr(M_2|D)}{\Pr(M_1|D)} = \frac{\Pr(M_2)}{\Pr(M_1)} \times \frac{\mathcal{Z}_{M_2}}{\mathcal{Z}_{M_1}} .
\end{equation}
Assuming there is no prior preference for one model or the other, i.e., $\Pr(M_1)=\Pr(M_2)$, the evidence ratio $\mathcal{Z}_{M_2}/\mathcal{Z}_{M_1}$ (often referred to as the Bayes' factor), or log-evidence difference~$\Delta\ln\mathcal{Z}$, becomes decisive.%
\footnote{\label{fn:subscripts}Note that in \cref{sec:methods_model_comparison} we are using the models $M_1$ and $M_2$ in the subscript of the evidence~$\mathcal{Z}$, since these are the compared entity, whereas the data~$D$ stay the same. This reverses in \cref{sec:methods_data_consistency}, where we emphasise the use of different data sets $A$ and $B$ as subscripts of likelihood~$\mathcal{L}$ and evidence~$\mathcal{Z}$, while the theoretical model remains fixed.}

Therefore the \cref{eq:occam_equation_special} plays a central role in model comparison, separating the log-evidence into an average fit and an Occam penalty. A more complex model (or extension) is favored only if it sufficiently improves the likelihood to offset the extra compression it imposes. Importantly, only \emph{constrained} additional parameters are penalized; hence, unconstrained extensions to a model leave $\ln\mathcal{Z}$ unchanged, as desired.%
\footnote{This is an important and desired effect, because otherwise we would, for example, end up with CMB data (which constrain all cosmological parameters) preferring the \LCDM model, but BAO or SN data wanting a model with fewer parameters, since they cannot constrain the optical depth~$\tau_\mathrm{reio}$ or the primordial parameters~$A_\mathrm{s}$ and~$n_\mathrm{s}$. We would then end up with strong disagreements in model preference between different data sets.}

\subsubsection{Pillar 3:\texorpdfstring{\\}{ }Data-set consistency and tension quantification}
\label{sec:methods_data_consistency}

Over the last couple of decades, the precision on cosmological parameters has increased to a level where we now need to verify the consistency of data from different probes first, before combining them to leverage degeneracy-breaking combinations of data sets.
To that end, there have been various suggestions for assessing concordance or discordance of data sets (for an early overview see, e.g., Ref.~\cite{Charnock2017}).

We will follow the propositions in Ref.~\cite{Handley2019_tension}, which are most aligned with the procedure that we outlined in the preceding sections. 
As a first conservative statistic we will look at the evidence consistency ratio~$\mathcal{R}$ between the joint evidence ${\mathcal{Z}_{AB}=\avp{\mathcal{L}_{AB}}}$ (with ${\mathcal{L}_{AB}=\mathcal{L}_A\mathcal{L}_B}$) and the individual evidence $\mathcal{Z}_A$ and $\mathcal{Z}_B$ for two data sets $A$ and $B$:
\begin{align}
    \label{eq:R}
    \mathcal{R} 
    &= \frac{\Pr(A \cup B|\mathsf{h_0},M)}{\Pr(A \cup B|\mathsf{h_1},M)} = \frac{\Pr(A \cup B|M)}{\Pr(A|M)\Pr(B|M)} \nonumber\\
    &= \frac{\mathcal{Z}_{AB}}{\mathcal{Z}_A\mathcal{Z}_B}, 
\end{align}
\begin{equation}
    \ln\mathcal{R} = \ln\mathcal{Z}_{AB} - \ln\mathcal{Z}_{A} - \ln\mathcal{Z}_{B},
\end{equation}
which quantifies the amount by which the hypothesis~$\mathsf{h_0}$ that the two data sets can be described by the \emph{same} set of parameters in a joint analysis compares to the alternative (extreme) hypothesis~$\mathsf{h_1}$ that these two data sets need to be described by completely \emph{distinct} parameters~\cite{Marshall2006,Raveri2016,Raveri2019}. A value $\ln\mathcal{R}\ll0$ signals tension, while a value $\ln\mathcal{R}\gg0$ indicates agreement between the data sets.\footnoteref{fn:subscripts}

Because of a potentially significant prior dependence of the evidence consistency ratio~$\mathcal{R}$, it has been criticized as being biased towards agreement of data sets~\cite{Raveri2019}.
However, as already cautiously indicated, the comparison between the hypotheses~$\mathsf{h_0}$ and $\mathsf{h_1}$ is somewhat extreme. Typically, the disagreement between two (or more) data sets only occurs in a subset of the full parameter space. Hence, the de-facto doubling of parameters and thereby a doubling of the Occam penalty correctly pushes toward a joint analysis, when e.g., the majority of parameters are in perfect agreement and only a single or a small subset of parameters are in tension. As such, the evidence ratio should be interpreted as a conservative measure that \textit{might} indicate agreement between data sets where there is none, but not the other way around. Therefore, any disagreement under $\mathcal{R}$ should be taken seriously.

Using the \cref{eq:occam_equation_special}, we can split the evidence in \cref{eq:R}, and thus $\mathcal{R}$ itself, into a goodness-of-fit part,
\begin{equation}
    \label{eq:S}
    \ln\mathcal{S} = \avP{\ln\mathcal{L}_{AB}} - \avP{\ln\mathcal{L}_A} - \avP{\ln\mathcal{L}_B},
\end{equation}
and an Occam penalty part,
\begin{equation}
    \hat{\mathcal{I}} = \mathcal{D}_A + \mathcal{D}_B - \mathcal{D}_{AB} .
\end{equation}
The likelihood consistency ratio~$\mathcal{S}$ was introduced as a prior-independent, more fit-driven alternative measure of tension to $\mathcal{R}$, and is also referred to as the ``suspiciousness''~\cite{Handley2019_tension}.
For a multivariate Gaussian posterior, $\tilde{\mathcal{d}}-2\ln\mathcal{S}$ is $\chi_{\tilde{\mathcal{d}}}^2$ distributed, allowing us to map the suspiciousness values to a $p$-value and a corresponding ``sigma value'' to quantify the level of tension (for a more detailed derivation, refer to Ref.~\cite{Handley2019_tension}):
\begin{align}
    \label{eq:tension_sigma}
    p = \int_{\tilde{\mathcal{d}}-2\ln\mathcal{S}}^\infty \chi^2_{\tilde{\mathcal{d}}}(x) \dd{x}, && 
    \sigma = \sqrt{2}\Erfc^{-1}(p),
\end{align}
where $\Erfc^{-1}$ is the inverse complementary error function and $\tilde{\mathcal{d}}$ is the number of shared constrained parameters between the data sets quantified by the difference
\begin{equation}
    \tilde{\mathcal{d}} = \mathcal{d}_A + \mathcal{d}_B - \mathcal{d}_{AB},
\end{equation}
between the Bayesian model dimensionalities~$\mathcal{d}$ from \cref{eq:model_dimensionality} of the individual versus the joint data-set analyses. 
\Cref{eq:tension_sigma} is the basis we use when we are quoting tensions in terms of $p$-value or ``sigmas'' in the rest of this paper.

The suspiciousness statistic is preserved under Box--Cox transformations~\cite{BoxCox1964} making the tension quantification from \cref{eq:tension_sigma} generalize to correlated or non-Gaussian cases, provided the underlying prior distribution does not become significantly distorted under the transformation in the region of the posterior bulk~\cite{Handley2019_tension,Lemos2020}.

Both the evidence ratio~$\mathcal{R}$ and the suspiciousness~$\mathcal{S}$ can be straight-forwardly generalized to more data sets using
\begin{alignat}{2}
    \mathcal{R}_{A,B,\ldots,N} &\equiv \frac{\mathcal{Z}_{A,B,\ldots,N}}{\mathcal{Z}_A\mathcal{Z}_B\cdots\mathcal{Z}_N}, \\
    \label{eq:multi_tension}
    \ln\mathcal{S}_{A,B,\ldots,N} 
    &= \avP{\ln\mathcal{L}_{AB\ldots N}} 
     &&- \avP{\ln\mathcal{L}_A} \nonumber\\
    &&&- \avP{\ln\mathcal{L}_B} \nonumber\\
    &&&  \cdots \nonumber\\
    &&&- \avP{\ln\mathcal{L}_N}.
\end{alignat}
Note that for $\mathcal{R}_{A,B,\ldots,N}$ the prior sensitivity gets stronger with every additional data set in the comparison. With the prior sensitivity removed, this is not a problem for $\ln\mathcal{S}_{A,B,\ldots,N}$. Hence, for the evidence-based tension statistic~$\mathcal{R}$ we choose to reduce this to a two-way comparison through initial combination of data sets.

\subsection{Theory codes and cosmological models}
\label{sec:models}

\begin{table*}[t]
    \centering
    \begin{tabularx}{1.0\textwidth}{l c c X}
        \toprule
            Parameter & Fixed Value & Prior Range & Description \\
        \midrule
            $\omega_\mathrm{b} \equiv \Omega_\mathrm{b}h^2$ &      & $0.019<\omega_\mathrm{b}<0.025$         & Baryon density today.                                                          \\
            $\omega_\mathrm{c} \equiv \Omega_\mathrm{c}h^2$ &      & $0.08<\omega_\mathrm{c}<0.3$            & Cold dark matter density today.                                                \\
            $h$                                             &      & $0.4<h<0.9$                             & Hubble parameter with ${H_0 \equiv 100\,h\,\unit{\km\per\s\per\mega\parsec}}$. \\
            $\tau_\mathrm{reio}$                            &      & $0.01<\tau_\mathrm{reio}<0.2$           & Optical depth due to reionization.                                             \\
            $A_\mathrm{s}$                                  &      & $2.6 < \ln(10^{10} A_\mathrm{s}) < 3.5$ & Amplitude of the scalar primordial power spectrum.                        \\
            $n_\mathrm{s}$                                  &      & $0.9<n_\mathrm{s}<1.04$                 & Spectral index or tilt of the scalar primordial power spectrum.                \\
        \midrule
            $M_\nu \equiv \sum{m_\nu}$                      & 0.06 & $0<M_\nu<1$                             & Sum of the neutrino masses (in \unit{\eV}); for \LCDM, we assume a single massive neutrino such that $M_\nu\!=\!m_\nu\!=\!\SI{0.06}{\eV}$, while for the neutrino extension, we assume 3 degenerate neutrino masses, i.e., $M_\nu\!=\!3m_\nu$. \\
            $\Omega_K$                                      & 0    & $-0.15 < \Omega_K < +0.15$              & Spatial curvature density parameter today.                                     \\
            $w$ or $w_0$                                    & $-1$ & $-2<w<0$                                & Constant equation-of-state parameter of dark energy.                           \\
            $w_a$                                           & 0    & $-3<w_a<+2$                             & Parameter for the time-varying part of the  equation-of-state parameter of dark energy, 
                                                                                                        with additional constraint $w(a=0)={w_0+w_a<0}$ to ensure dark energy can eventually dominate.\\
        \bottomrule
    \end{tabularx}
    \caption{\label{tab:priors}
        Prior ranges of the cosmological sampling parameters, assuming uniform sampling in the specified range. The second block shows the parameters that are fixed in the baseline $\Lambda$CDM model with the values specified in the second column, but sampled in the minimal extensions of $\Lambda$CDM listed in \cref{sec:models}.
    }
\end{table*}

In this paper, we use the standard \LCDM model as our baseline, and look at four minimal extensions.
\begin{description}
    \item[\nuLCDM] The single massive neutrino with a fixed mass of $m_\nu=\SI{0.06}{\eV}$ in the \LCDM base model is changed to sampling $M_\nu\equiv\sum{m_\nu}$, the sum of three degenerate massive neutrinos.
    \item[\KLCDM] The curvature density parameter~\OmegaK is sampled symmetrically around the zero value assumed in the base \LCDM model. 
    \item[\wCDM] The cosmological constant~$\Lambda$ is replaced by a variable (but constant in time) equation-of-state parameter of dark energy~$w$.
    \item[\wwaCDM] The cosmological constant~$\Lambda$ is replaced by evolving dark energy parameterized by $w_0$ and $w_a$. with a resulting equation-of-state parameter of dark energy of $w(a)=w_0+w_a(1-a)$, where $a$ is the scale factor.
\end{description}
The prior ranges of the cosmological parameters can be found in \cref{tab:priors}.

We have generated emulators for each of these five cosmological models using \texttt{CosmoPower}~\cite{Mancini2022}.\footnote{\url{https://github.com/alessiospuriomancini/cosmopower}} For the training we created random Latin hypercubes with \num{200000} samples of the cosmological parameter sets for each model, and then used the Boltzmann code \texttt{CLASS}~\cite{Class1_Lesgourgues2011,Class2_Blas2011,Class4_neutrinos_Lesgourgues2011,Class_OmegaK_Lesgourgues2014}\footnote{\url{https://github.com/lesgourg/class_public}} to compute the CMB angular power spectra for temperature, polarization, and lensing in the multipole range $2 \leq \ell \leq 5000$, and the evolution of the Hubble parameter~$H(z)$ and the angular diameter distance~$D_\mathrm{A}(z)$ in the redshift range $0 \leq z \leq 20$, with a step size of $\Delta{z}=0.005~$.

We then interfaced our \texttt{CosmoPower} emulators with \texttt{Cobaya}~\cite{Cobaya_Torrado2021},\footnote{\url{https://github.com/CobayaSampler/cobaya}} which we used to sample various likelihoods (see \cref{sec:data,tab:likelihood_overview}) with the adaptive Markov chain Monte Carlo~(MCMC) sampler based on \texttt{CosmoMC}~\cite{CosmoMC_Lewis2002,CosmoMC_Lewis2013}, and the nested sampler \texttt{PolyChord}~\cite{PolyChord1,PolyChord2},\footnote{\url{https://github.com/PolyChord/PolyChordLite}} tailored for high-dimensional evidence computation. The post-processing and visualization of the MCMC and nested sampling chains was done with \texttt{anesthetic}~\cite{Anesthetic}.\footnote{\url{https://github.com/handley-lab/anesthetic}}

\subsection{Data}
\label{sec:data}

\begin{table*}[p]
    \centering
    \begin{tabularx}{\textwidth}{l l c l S >{\centering\arraybackslash}X l}
\toprule
Label & Data set & Tracer & \texttt{Cobaya} Likelihood & $z_\mathrm{eff}$ & Range & Observables \\
\midrule
(SimAll +)
& Planck PR3 & low-$\ell~TT$ & \texttt{planck\_2018\_lowl.TT\_clik} \cite{Planck2018_likelihoods}                                              & & $2\leq\ell\leq30$    & $C_\ell^{TT}$ \\\addlinespace[\medskipamount] \textbf{Plik}
& Planck PR3 & low-$\ell~EE$ & \texttt{planck\_2018\_lowl.EE\_clik} \cite{Planck2018_likelihoods}                                              & & $2\leq\ell\leq30$    & $C_\ell^{EE}$ \\\addlinespace[\medskipamount]
& Planck PR3 & $TTTEEE$      & \texttt{planck\_2018\_highl\_plik.TTTEEE} \cite{Planck2018_likelihoods}                                         & & $30\leq\ell\leq2500$ & $C_\ell^{TT}$, $C_\ell^{TE}$, $C_\ell^{EE}$ \\
\midrule
(SimAll +)
& Planck PR3 & low-$\ell~TT$ & \texttt{planck\_2018\_lowl.TT\_clik} \cite{Planck2018_likelihoods}                                              & & $2\leq\ell\leq30$    & $C_\ell^{TT}$ \\\addlinespace[\medskipamount] \textbf{CamSpec}
& Planck PR3 & low-$\ell~EE$ & \texttt{planck\_2018\_lowl.EE\_clik} \cite{Planck2018_likelihoods}                                              & & $2\leq\ell\leq30$    & $C_\ell^{EE}$ \\\addlinespace[\medskipamount]
& Planck PR4 & $TTTEEE$      & \multicolumn{2}{l}{\texttt{planck\_NPIPE\_highl\_CamSpec.TTTEEE} \cite{CamSpec4_Rosenberg2022}}                   & $30\leq\ell\leq2500$ & $C_\ell^{TT}$, $C_\ell^{TE}$, $C_\ell^{EE}$ \\
\midrule
(Lollipop +)
& Planck PR3 & low-$\ell~TT$ & \texttt{planck\_2018\_lowl.TT} \cite{Planck2018_likelihoods}                                                    & & $2\leq\ell\leq30$    & $C_\ell^{TT}$ \\\addlinespace[\medskipamount] \textbf{Hillipop}
& Planck PR4 & low-$\ell~EE$ & \texttt{planck\_2020\_lollipop.lowlE} \cite{Lollipop_Hamimeche2008,Lollipop_Mangilli2015,Lollipop_Tristram2021} & & $2\leq\ell\leq30$    & $C_\ell^{EE}$ \\\addlinespace[\medskipamount]
& Planck PR4 & $TTTEEE$      & \multicolumn{2}{l}{\texttt{planck\_2020\_hillipop.TTTEEE\_bin} \cite{Hillipop_Tristram2023}}                     & $30\leq\ell\leq2500$ & $C_\ell^{TT}$, $C_\ell^{TE}$, $C_\ell^{EE}$ \\
\midrule
\textbf{L3}
& Planck PR3 & CMB lensing   & \texttt{planck\_2018\_lensing.clik} \cite{Planck2018_lensing}  & & $8 \leq L \leq 400$  & $C_L^{\phi\phi}$ \\
\midrule
\textbf{L4}
& Planck PR4 & CMB lensing   & \texttt{PlanckPR4Lensing} \cite{Carron2022}                     & & $8 \leq L \leq 400$  & $C_L^{\phi\phi}$ \\
\midrule
\textbf{SDSS\,12}
& 6dFGS      &     & \texttt{bao.bao.sixdf\_2011\_bao} \cite{Beutler2011,Beutler2012}            & 0.106 &              & $r_\mathrm{d}/D_V$                     \\\addlinespace[\medskipamount]
& SDSS DR7   & MGS & \texttt{bao.sdss\_dr7\_mgs} \cite{SDSS7_Ross2015}                           & 0.15  & $0.07<z<0.2$ & $D_V/r_\mathrm{d}$                     \\\addlinespace[\medskipamount]
& SDSS DR12  & LRG & \texttt{bao.sdss\_dr12\_consensus\_bao} \cite{SDSS12_2017}                  & 0.38  & $0.2<z<0.5$  & $D_M/r_\mathrm{d}$, $r_\mathrm{d}H$    \\
& SDSS DR12  & LRG & \texttt{bao.sdss\_dr12\_consensus\_bao} \cite{SDSS12_2017}                  & 0.51  & $0.4<z<0.6$  & $D_M/r_\mathrm{d}$, $r_\mathrm{d}H$    \\
& SDSS DR12  & LRG & \texttt{bao.sdss\_dr12\_consensus\_bao} \cite{SDSS12_2017}                  & 0.61  & $0.5<z<0.75$ & $D_M/r_\mathrm{d}$, $r_\mathrm{d}H$    \\
\midrule
\textbf{SDSS\,16}
& SDSS DR7   & MGS & \texttt{bao.sdss\_dr7\_mgs} \cite{SDSS7_Ross2015}                           & 0.15  & $0.07<z<0.2$ & $D_V/r_\mathrm{d}$                     \\\addlinespace[\medskipamount]
& SDSS DR12  & LRG & \texttt{bao.sdss\_dr12\_lrg\_bao\_dmdh} \cite{SDSS12_2017}                  & 0.38  & $0.2<z<0.5$  & $D_M/r_\mathrm{d}$, $D_H/r_\mathrm{d}$ \\
& SDSS DR12  & LRG & \texttt{bao.sdss\_dr12\_lrg\_bao\_dmdh} \cite{SDSS12_2017}                  & 0.51  & $0.4<z<0.6$  & $D_M/r_\mathrm{d}$, $D_H/r_\mathrm{d}$ \\
& SDSS DR16  & LRG & \texttt{bao.sdss\_dr16\_lrg\_bao\_dmdh} \cite{SDSS16_2021}                  & 0.698 & $0.6<z<1.0$  & $D_M/r_\mathrm{d}$, $D_H/r_\mathrm{d}$ \\\addlinespace[\medskipamount]
& SDSS DR16  & ELG & \texttt{bao.sdss\_dr16\_bao\_elg} \cite{SDSS16_2021}                        & 0.845 & $0.6<z<1.1$  & $D_V/r_\mathrm{d}$                     \\\addlinespace[\medskipamount]
& SDSS DR16  & QSO & \texttt{bao.sdss\_dr16\_qso\_bao\_dmdh} \cite{SDSS16_2021}                  & 1.48  & $0.8<z<2.2$  & $D_M/r_\mathrm{d}$, $D_H/r_\mathrm{d}$ \\\addlinespace[\medskipamount]
& SDSS DR16  & Ly$\alpha\times$QSO & \texttt{bao.sdss\_dr16\_baoplus\_lyxqso} \cite{SDSS16_2021} & 2.334 & $1.77<z$ & $D_M/r_\mathrm{d}$, $D_H/r_\mathrm{d}$     \\\addlinespace[\medskipamount]
& SDSS DR16  & Ly$\alpha$          & \texttt{bao.sdss\_dr16\_baoplus\_lyauto} \cite{SDSS16_2021} & 2.334 & $2.1<z$  & $D_M/r_\mathrm{d}$, $D_H/r_\mathrm{d}$     \\
\midrule
\textbf{DESI\,2}
& DESI DR2  & BGS        & \texttt{bao.desi\_dr2.desi\_bao\_bgs}                                 & 0.295 & $0.1<z<0.4$ & $D_V/r_\mathrm{d}$ \\\addlinespace[\medskipamount] \cite{DESI2_2025_cosmology}
& DESI DR2  & LRG        & \texttt{bao.desi\_dr2.desi\_bao\_lrg1}                                & 0.510 & $0.4<z<0.6$ & $D_M/r_\mathrm{d}$, $D_H/r_\mathrm{d}$ \\
& DESI DR2  & LRG        & \texttt{bao.desi\_dr2.desi\_bao\_lrg2}                                & 0.706 & $0.6<z<0.8$ & $D_M/r_\mathrm{d}$, $D_H/r_\mathrm{d}$ \\
& DESI DR2  & LRG \& ELG & \texttt{bao.desi\_dr2.desi\_bao\_lrg3pluselg1}                        & 0.934 & $0.8<z<1.1$ & $D_M/r_\mathrm{d}$, $D_H/r_\mathrm{d}$ \\
& DESI DR2  & ELG        & \texttt{bao.desi\_dr2.desi\_bao\_elg2}                                & 1.321 & $1.1<z<1.6$ & $D_M/r_\mathrm{d}$, $D_H/r_\mathrm{d}$ \\\addlinespace[\medskipamount]
& DESI DR2  & QSO        & \texttt{bao.desi\_dr2.desi\_bao\_qso}                                 & 1.484 & $0.8<z<2.1$ & $D_M/r_\mathrm{d}$, $D_H/r_\mathrm{d}$ \\\addlinespace[\medskipamount]
& DESI DR2  & Ly$\alpha$ & \texttt{bao.desi\_dr2.desi\_bao\_lya}                                 & 2.330 & $1.8<z<4.2$ & $D_M/r_\mathrm{d}$, $D_H/r_\mathrm{d}$ \\
\midrule
\textbf{DES\,y5}
& DES year 5 & SN Ia      & \texttt{sn.desy5} \cite{DESy5_2024}                 & \multicolumn{2}{c}{$0.02509<z<1.12132$} & SN distance \\
\midrule
\textbf{Pantheon$^+$}
& Pantheon+ & SN Ia   & \texttt{sn.pantheonplus} \cite{PantheonPlus_Brout2022} & \multicolumn{2}{c}{$0.008<z<2.26$}      & SN distance \\
\midrule
\textbf{Union3}
& Union3    & SN Ia      & \texttt{sn.union3} \cite{Union3_Rubin2025}          & \multicolumn{2}{c}{$0.05<z<2.26226$}    & SN distance \\
\midrule
\multicolumn{2}{l}{\textbf{Freedmann 2020}}
            & TRGB       & \texttt{H0.freedman2020} \cite{Freedman2020}        &       &             &                    \\
\midrule
\multicolumn{2}{l}{\textbf{Riess 2020}}
            & Cepheids   & \texttt{H0.riess2020} \cite{Riess2020}              &       &             &                    \\
\bottomrule
    \end{tabularx}
    \caption{
        Likelihoods used in this paper. 
        The first column shows the label we use to refer to the likelihoods in the corresponding block. Unless otherwise specified, we only use the names of the high-$\ell$ likelihoods in our CMB labels, with the low-$\ell$ likelihoods implied.
        The second column specifies which data release the corresponding likelihood is coming from.
        Note that we are not using any redshift-space distortion~(RSD) data. We only use the baryon acoustic oscillation~(BAO) data from the 6-degree Field Galaxy Survey~(6dFGS), the Sloan Digital Sky Survey~(SDSS), and the Dark Energy Spectroscopic Instrument~(DESI).
    }
    \label{tab:likelihood_overview}
\end{table*}

\Cref{tab:likelihood_overview} provides an overview of the likelihoods used for this paper, and introduces the labels we use for groupings of these likelihoods.

\subsubsection{Cosmic microwave background (CMB)}
\label{sec:data_cmb}

As the only cosmological probe able to constrain all six cosmological parameters jointly, the cosmic microwave background~(CMB) forms the backbone for our analysis.
We compare two releases of \Planck data, namely the \Planck public release~3~(PR3) and \Planck public release~4~(PR4).
Planck~PR3 is also referred to as the ``\Planck~2018'' or ``legacy'' release~\cite{Planck2018_overview}.
Planck~PR4 is also referred to under the name ``NPIPE'' after the processing pipeline that was used in the joint re-analysis of the \Planck Low-Frequency Instrument~(LFI) and High-Frequency Instrument~(HFI) data, including additional repointing data and improved detector calibration~\cite{NPIPE}.

The CMB data are split into three likelihoods, two built from low-multipole ($\ell\leq30$) angular power spectra, one for temperature (low\=/$\ell$ $TT$) and the other for polarization $E$ modes (low\=/$\ell$ $EE$), and one built from high-multipole ($30<\ell\leq2500$) cross-frequency spectra for temperature and $E$-mode polarization (high\=/$\ell$ $TTTEEE$). Additionally, both PR3 and PR4 come with independent CMB lensing likelihoods, which we refer to as ``L3'' \cite{Planck2018_lensing} and ``L4'' \cite{Carron2022}.

For the low-multipole temperature data, we use the \Commander~\cite{Planck2018_likelihoods}\footnote{\label{fn:PLA}\url{https://pla.esac.esa.int/\#cosmology}} likelihood derived from Gibbs-sampled component-separated PR3 maps. We use the \Commander likelihood for all CMB runs, unless we specifically refer to high\=/$\ell$ only.

For the low-multipole $E$-mode polarization data, we use either the \Planck PR3 \SimAll~\cite{Planck2018_likelihoods}\footnoteref{fn:PLA} likelihood, which compares cross-spectra between the \SIlist{100;143}{\GHz} HFI channels to simulations, or we use the \Lollipop~\cite{Lollipop_Mangilli2015,Lollipop_Tristram2021}\footnote{\url{https://github.com/planck-npipe/lollipop/}} likelihood, which implements a Hamimeche \& Lewis approximation~\cite{Lollipop_Hamimeche2008} based on the NPIPE component-separated polarization maps. As for the low\=/$\ell$ $TT$ likelihood, we use these low\=/$\ell$ $EE$ likelihoods for all CMB runs, unless specifically labeling with ``high\=/$\ell$'' only.

Finally, for the high-multipole data we compare between the \Plik~\cite{Planck2018_likelihoods},\footnoteref{fn:PLA} \CamSpec~\cite{CamSpec4_Rosenberg2022},\footnote{\url{https://github.com/CobayaSampler/cobaya/tree/master/cobaya/likelihoods/planck_NPIPE_highl_CamSpec/}} and \Hillipop~\cite{Hillipop_Tristram2023}\footnote{\url{https://github.com/planck-npipe/hillipop/}} likelihoods.
All three are built on multiple cross-frequency spectra between the \SIlist{100;143;217}{\GHz} maps with nuisance parameters for astrophysical foregrounds and instrumental systematics.
\Plik uses half-mission PR3 maps to construct four cross-spectra (${100\!\times\!100}$, ${143\!\times\!143}$, ${217\!\times\!217}$, ${143\!\times\!217}$) for $TT$ and all six cross-spectra combinations for $TE$ and $EE$, and comes with 20 nuisance parameters (two for frequency calibrations, ten for Galactic and eight for extragalactic foregrounds), in addition to the shared overall calibration parameter~$A_\mathrm{Planck}$ with the low\=/$\ell$ likelihoods.
Both \CamSpec and \Hillipop use cross-spectra from PR4 ``detset'' maps of the \SIlist{100;143;217}{\GHz} channels, i.e., the data are split across detector sets for each frequency rather than in the time domain, improving the noise handling. 
\CamSpec does not include the \SI{100}{\GHz} channel in the $TT$ cross-spectra, but uses all six detset cross-spectra combinations for $TE$ and $EE$. 
It precedes with an initial galactic dust-cleaning step using \Planck's \SI{353}{\GHz} channel, after which it samples with the overall calibration parameter $A_\mathrm{Planck}$ and with eight further nuisance parameters: two for overall $TE$ and $EE$ polarization calibrations; and six describing the angular power spectra of all foreground residuals in $TT$ combined, in the form of power laws (i.e., with amplitudes~$A_{\nu_1\times\nu_2}^\mathrm{power}$ and spectral indices~$\gamma_{\nu_1\times\nu_2}^\mathrm{power}$ for the cross-frequencies ${143\!\times\!143}$, ${217\!\times\!217}$, ${143\!\times\!217}$ in \unit{\GHz}).
\Hillipop uses all six detset cross-spectra combinations for the $TT$, $TE$, and $EE$ spectra, and does not involve any pre-cleaning steps.
Instead, it comes with 16 sampled nuisance parameters in addition to $A_\mathrm{Planck}$: five for the cross-frequency calibrations; four for Galactic foregrounds; and seven for extragalactic foregrounds (cosmic infrared background, thermal and kinetic Sunyaev--Zeldovich effects, and point sources).

When combining low\=/$\ell$ and high\=/$\ell$ likelihoods, we adopt the same combinations as used in the published high\=/$\ell$ likelihood analyses of Refs.~\cite{Planck2018_likelihoods,CamSpec4_Rosenberg2022,Hillipop_Tristram2023}.
Both the \Planck~PR3 and the \Planck~PR4 CMB lensing likelihoods use a minimum variance quadratic estimator~\cite{Planck2018_lensing,Carron2022}.\footnoteref{fn:PLA}$^,$\footnote{\url{https://github.com/carronj/planck_PR4_lensing}}

Note that over the last several months the 6th data release~(DR6) of the Atacama Cosmology Telescope~(ACT) and the first data release~(DR1) of the 3rd generation South Pole Telescope~(SPT-3G) were published, which are now approaching the constraining power of the \Planck data, and are more sensitive for high multipoles~\cite{ACT6_Louis2025,SPT3G_Camphuis2025}. Regarding the comparisons of the added value of these data sets for the topics explored in this paper, beyond the analysis presented in Ref.~\cite{Tristram2025_planck_act_spt}, we defer detailed investigation to future work.

\subsubsection{Baryon acoustic oscillations (BAOs)}
\label{sec:data_bao}

For baryon acoustic oscillations~(BAOs) we group the data into three generations of surveys. 
The set that was available at the time of \Planck~PR3 used single distance estimates from the 6-degree Field Galaxy Survey~(6dFGS)~\cite{Beutler2011,Beutler2012} and the main galaxy sample~(MGS) provided by SDSS data release~(DR)~7~\cite{SDSS7_Ross2015}, along with the three distance estimates from luminous red galaxies~(LRGs) provided by SDSS~DR12~\cite{SDSS12_2017}.

This was later updated with the SDSS~DR16~\cite{SDSS16_2021}, which additionally used estimates from emission-line galaxies~(ELGs), quasars~(QSO), the Lyman-$\alpha$ forest~(Ly$\alpha$), and the cross-correlation of quasars with Lyman-$\alpha$. 

More recently DESI released results from their first~(DR1) and second~(DR2) data releases of BAO measurements in quick succession~\cite{DESI1_2024_cosmology,DESI2_2025_cosmology}, updating the full redshift range previously jointly covered by SDSS DR7, DR12, and DR16. They kept the same naming convention for the tracers in the redshift bins, except for the first, which they renamed the ``bright galaxy sample''~(BGS).
The relevant likelihoods for the BAO data are listed in Table~\ref{tab:likelihood_overview}.

\subsubsection{Type Ia supernovae (SNe)}
\label{sec:data_sn}

We contrast two type~Ia supernovae~(SNe) data sets in our analysis.
The Dark Energy Survey year-5~(DESy5) likelihood contains 1829 supernovae in a redshift range of about $0<z<1.13$~\cite{DESy5_2024}.
The \PantheonPlus~(P$^+$) likelihood contains 1701 supernovae in a redshift range of about $0<z<2.3$~\cite{PantheonPlus_Brout2022}.

\subsubsection{Distance-ladder data}

While not the focus of this work, we include the results from two local measurements of the Hubble parameter~$H_0$, to provide a point of comparison to the much discussed ``Hubble tension''.
We use Gaussian likelihoods based on the results from distance-ladder calibration using either the tip of the red giant branch~(TRGB)~\cite{Freedman2020} or Cepheid variables~\cite{Riess2020}.

\section{Results}
\label{sec:results}

\begin{figure*}[p]
    \includegraphics[scale=0.99]{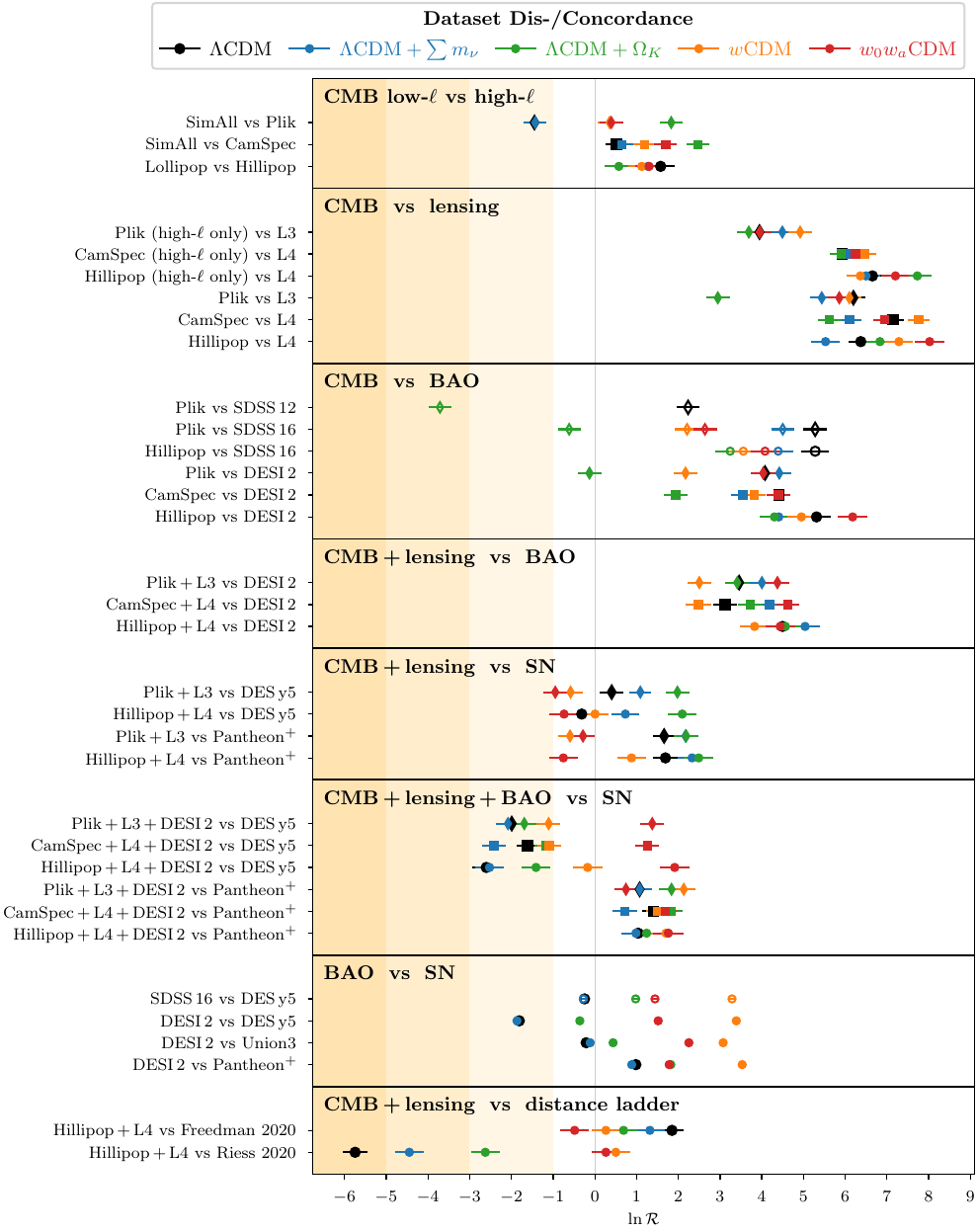}
    \caption{\label{fig:tension_R}
        Data-set tension based on the Bayesian evidence ratio statistic~$\mathcal{R}$ from \cref{eq:R} for a range of data-set combinations as labeled on the left-hand side.
        The black points correspond to the \LCDM model and the colored points to its extensions with neutrinos (blue), curvature (green), constant (orange), and dynamical (red) dark energy.
        Symbols with different shapes highlight different CMB likelihoods.
        The error bars reflect the sampling uncertainty.
        A value $\ln\mathcal{R}\ll0$ indicates tension, with the orange shaded regions referring to the Jeffrey's scale of moderate, strong, and very strong tension. $\ln\mathcal{R}\gg0$ indicates agreement.
    }
\end{figure*}

\begin{figure*}[p]
    \includegraphics[scale=0.99]{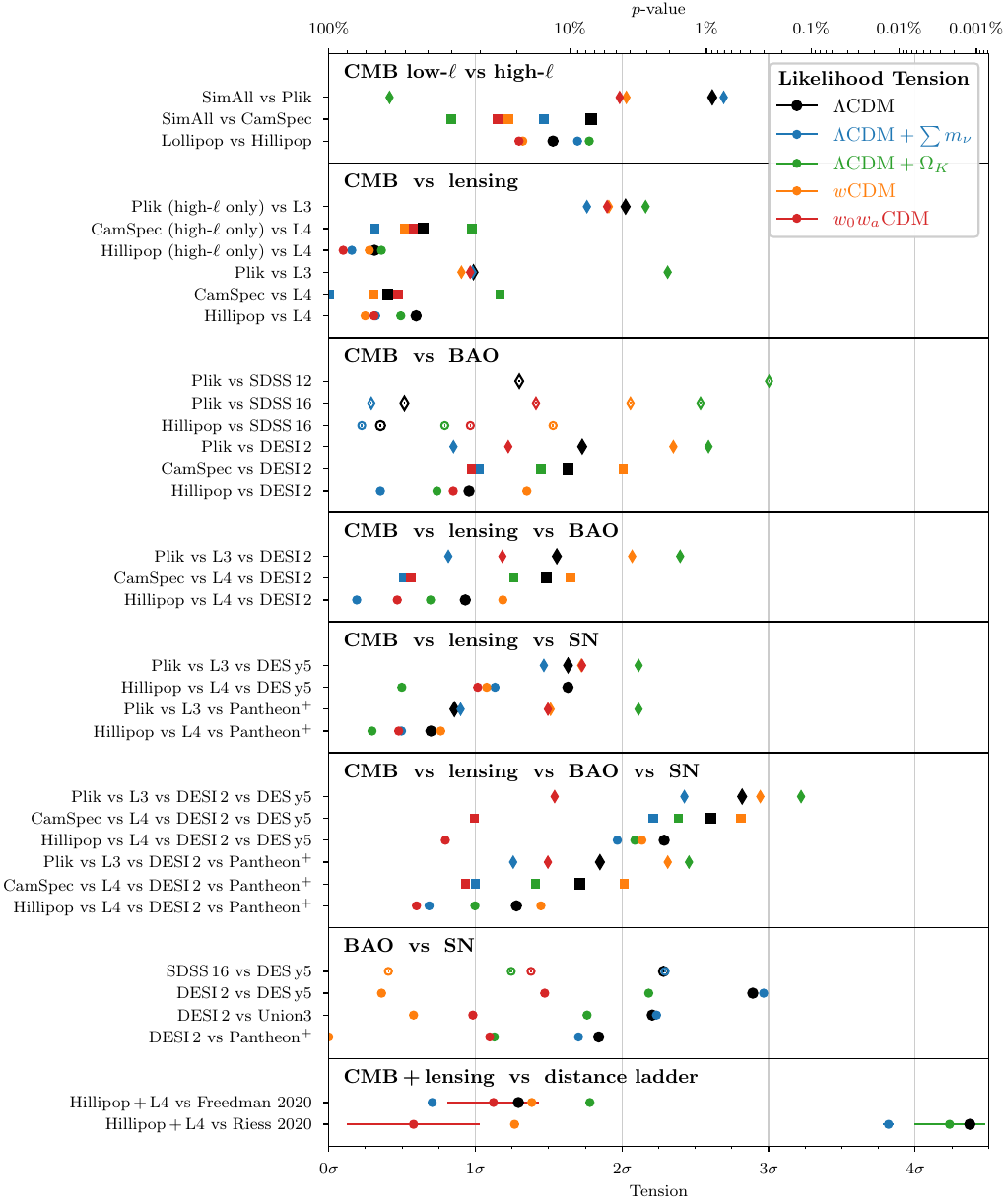}
    \caption{\label{fig:tension}
        Data-set tensions based on the average likelihood-ratio statistic~$\mathcal{S}$ from \cref{eq:S} for a range of data-set combinations as labeled on the left-hand side, expressed as a $p$-value (top axis) and numbers of $\sigma$ (bottom axis).
        Repeated ``vs'' implies 3-way or 4-way comparisons according to \cref{eq:multi_tension}.
        The black points correspond to the \LCDM model and the colored points to its extensions with neutrinos (blue), curvature (green), constant dark energy (orange), and dynamical dark energy (red).
        Symbols with different shapes highlight different CMB likelihoods.
        The error bars reflect the sampling uncertainty. Some error bars appear surprisingly big, which is an effect of the translation from $\mathcal{S}$ to a $p$-value using \cref{eq:tension_sigma} for a vanishing dimensionality $\tilde{\mathcal{d}}$, i.e., when $\mathcal{d}_A + \mathcal{d}_B \approx \mathcal{d}_{AB}$.
    }
\end{figure*}

\begin{figure*}[p]
    \includegraphics[scale=0.99]{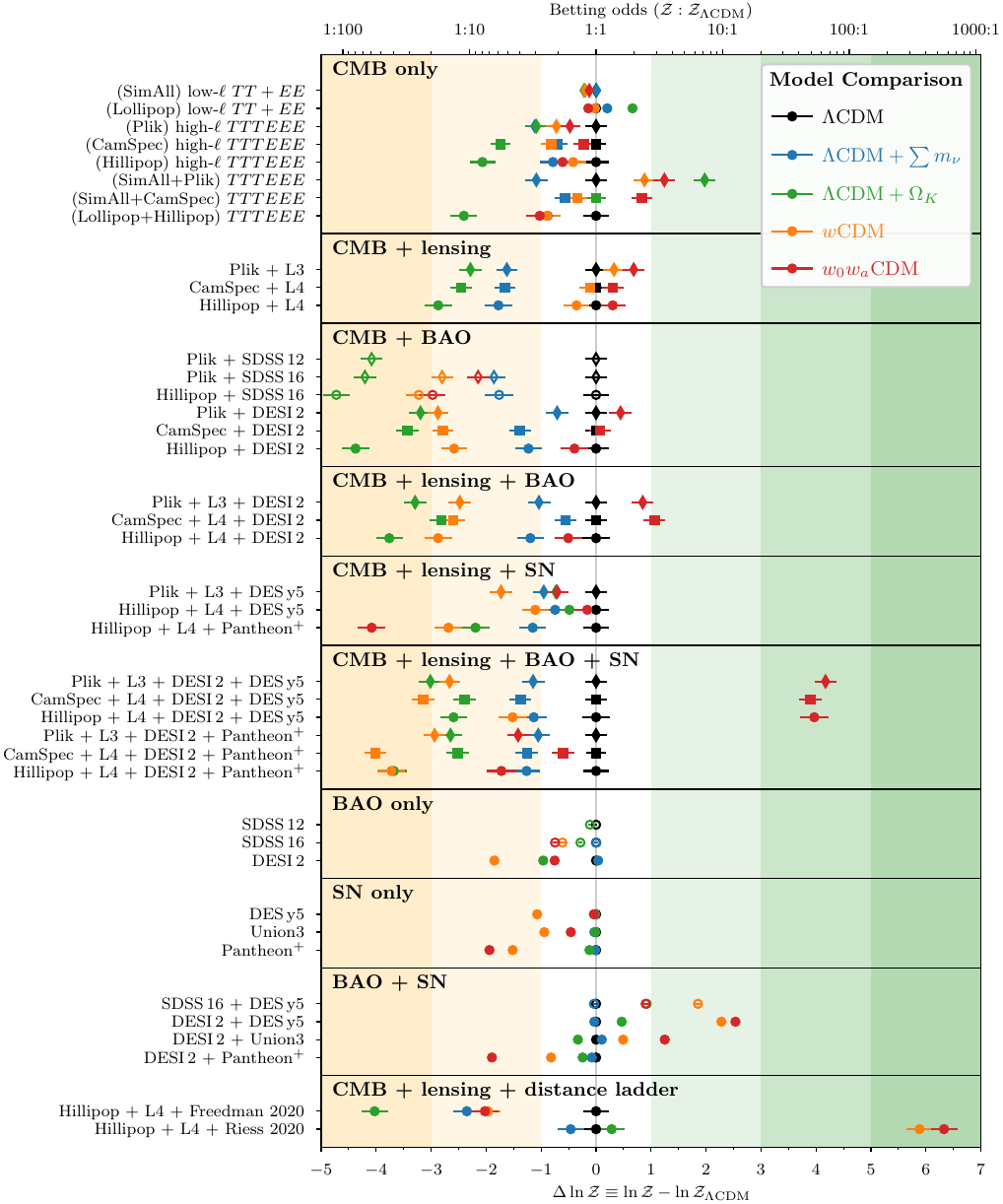}
    \caption{\label{fig:model_comparison}
        Model comparisons relative to \LCDM in terms of Bayesian evidence ratios (see \cref{eq:model_comparison}) based on different data-set combinations.
        The black points correspond to the \LCDM model and the colored points to its extensions with neutrinos (blue), curvature (green), constant dark energy (orange), and dynamical dark energy (red).
        The error bars reflect the sampling uncertainty.
        The shaded regions illustrate the Jeffreys' scale, categorizing the extensions as weakly, strongly, or very strongly favored compared to \LCDM in green, or disfavored in orange.
    }
\end{figure*}

\begin{figure}[tb]
    \includegraphics[width=\columnwidth]{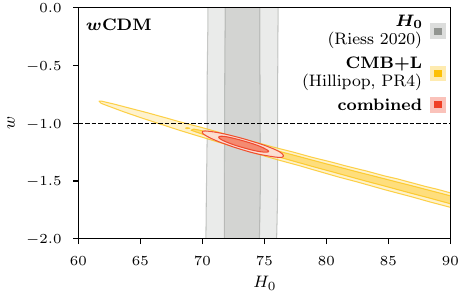}
    \caption{\label{fig:H0_w}
        Example of how increasing the parameter space from \LCDM (where $w=-1$) to \wCDM relieves the Hubble tension.
        However, this is not because both observables agree on $H_0$, but because the CMB loses almost all constraining power on $H_0$ when the parameter space is opened up.
    }
\end{figure}

\Cref{fig:tension_R,fig:tension,fig:model_comparison} lay out all the tension and model comparison results of the five cosmological models considered here (\LCDM, \nuLCDM, \KLCDM, \wCDM, \wwaCDM) for a large suite of likelihood combinations.
We show the $\ln\mathcal{R}$ statistic from \cref{eq:R} in \cref{fig:tension_R}, the tension from the $\ln\mathcal{S}$ statistic in numbers of $\sigma$ from \cref{eq:S,eq:tension_sigma} in \cref{fig:tension}, and the log-evidence difference $\Delta\ln\mathcal{Z}\equiv\ln\mathcal{Z}-\ln\mathcal{Z}_{\LCDM}$ relative to the \LCDM baseline in \cref{fig:model_comparison}.
These figures give a graphical way of representing consistency and inconsistency, derived within a unified and consistent framework, based on comparable estimators. Since the range of results may be a bit overwhelming at first glance, we will scrutinize the results in the rest of this section, as well as highlighting the most important findings.

This pictorial overview allows us to extract some broad perspectives.
First of all, based on \cref{fig:tension_R}, only very few data sets should be considered to be in tension, with only three points below $\ln\mathcal{R}=-3$.
Secondly, the same general conclusion is also reflected in \cref{fig:tension}, where only a handful of points exceed the $\SI{3}{\sigma}$ threshold.
And finally, \cref{fig:model_comparison} complements this picture, with the vast majority of points below the \LCDM normalization line. Only five data--model combinations result in a strong preference for an extension to \LCDM.

The data-set combination with the clearest tension corresponds to the widely discussed Hubble (or $H_0$) tension between CMB data from \Planck and local measurements of the Hubble parameter~$H_0$ involving calibration with Cepheid variables by Ref.~\cite{Riess2020}. This corresponds to the last row of points in \cref{fig:tension_R,fig:tension,fig:model_comparison}, and is included here mostly to help put all other tensions and model preferences for the various data--model combinations into perspective.

There are a few high-tension data combinations, where either the \wCDM or \wwaCDM models perform better than \LCDM, owing to the significant relaxation in constraining power that typically comes with the additional dark energy sampling parameters $w$, or $w_0$ and $w_a$. 
\Cref{fig:H0_w} shows this explicitly for the case of the $H_0$ tension. When switching from \LCDM to \wCDM, the CMB data from \Planck lose almost all their constraining power on the Hubble parameter and we are left with a strong degeneracy between $H_0$ and $w$. This degeneracy can then be broken by the local $H_0$ measurement.\footnote{Note that constraints from BAO data are not as degenerate in \wCDM as is the case for the CMB, so while moving to \wCDM might relieve the Hubble tension between CMB and local measurements, there would still remain a Hubble tension with BAOs.}

However, this type of easing of tensions is highly unsatisfying. Instead of changing the theory in a way that makes two data sets agree on a parameter, it effectively just increases the uncertainty of one data set to a level where it is no longer informative in that parameter.

In the following we will highlight in more detail a few points that stand out in \cref{fig:tension_R,fig:tension,fig:model_comparison}.

\subsection{CMB: internal consistency}
\label{sec:cmb_internal}

We already discussed how the $H_0$ tension helps us place various other tensions in perspective. Similarly it is instructive to look at \Planck internal consistencies to calibrate our tension scale. The first block in \cref{fig:tension_R,fig:tension,fig:model_comparison} compares CMB multipole splits, and the second block compares CMB with CMB lensing.

For the multipole splits, we combine the low\=/$\ell$ temperature and $E$-mode polarization likelihoods and compare them to the high\=/$\ell$ $TTTEEE$ likelihoods.
There are no points indicating any serious tension that would surpass a $\SI{3}{\sigma}$ threshold, but considering that sometimes even $\SI{2}{\sigma}$ ``tensions'' are quoted in the literature~\cite{Abdalla2022,Gu2025_desi_tension,Ye2025_desi_tension}, it is worth pointing out that the baseline cosmology of the \Planck~PR3 legacy release using the \Plik likelihood has a $\SI{2.6}{\sigma}$ tension between the low and the high multipoles (and see Ref.~\cite{Planck2017_IntLI} for further discussion of \Planck internal consistency).

However, there is a clear trend that updating from the \Plik to either the \CamSpec or \Hillipop likelihoods reduces this multipole tension, fairly independent of the cosmological model considered.
The curvature extension is the exceptional case where the low\=/$\ell$ versus high\=/$\ell$ split shows less tension under \Plik than \CamSpec or \Hillipop, which we discuss in more detail in \cref{sec:curvature}.

Turning to the second block in \cref{fig:tension_R,fig:tension}, we observe a similar picture. The high\=/$\ell$ \Plik likelihood is about $\SI{2}{\sigma}$ in tension with the PR3 lensing likelihood across all considered models, but switching to the PR4 CMB and lensing likelihoods reduces this to below $\SI{1}{\sigma}$. Combining \Plik with the low\=/$\ell$ \SimAll likelihood also reduces the tension to below $\SI{1}{\sigma}$ for all models (again with the exception of the curvature extension).
\CamSpec and \Hillipop are consistent with the PR4 lensing likelihood with and without inclusion of the low\=/$\ell$ likelihoods.

This approach enables the visualization and quantification of the differences arising from the choice of data and analysis methods for \Planck, as detailed in \cref{sec:data_cmb}. It further underscores the importance of discussing these differences in uncertainty in any analysis combining \Planck likelihoods with any other data sets.

\subsection{Curved universes ~ (Curvature tension?)}
\label{sec:curvature}

The curvature extension stands out wherever the \Plik likelihood is involved. When comparing low\=/$\ell$ to high\=/$\ell$, there is a surprisingly large gap in the tension statistic in \cref{fig:tension}, separating the curvature extension from \LCDM and all other extensions. This picture is inverted when comparing to additional observables (CMB lensing, BAOs, SNe), where now the curvature extension consistently exhibits the highest tension of all models considered.
On the model comparison side (\cref{fig:model_comparison}), it is curious how neither low\=/$\ell$ nor high\=/$\ell$ favor curvature on their own, but their combination shows a preference for curvature, which again vanishes with the inclusion of other observables.

\begin{figure}[tbp]
    \includegraphics[width=\columnwidth]{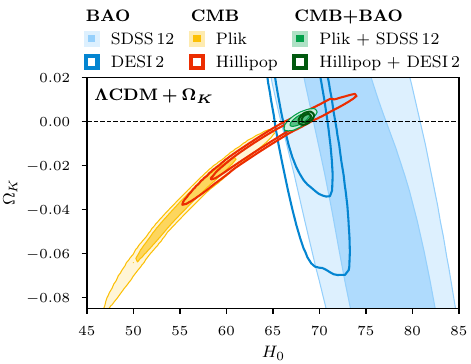}
    \caption{\label{fig:curvature_post_OmegaK_H0}
        \KLCDM posteriors for various data sets. Note the strong preference for closed universes under PR3, which gets significantly reduced when switching to PR4, making CMB data more compatible with the results from BAOs.
    }
\end{figure}

This peculiarity was already noticed at the time of the \Planck PR3 results, which with the latest available BAO data at that time (from SDSS DR12) led to a roughly $\SI{3}{\sigma}$ tension, referred to as the ``curvature tension''~\cite{DiValentino2020_curvature,Handley2021_curvature}.
This seemed particularly interesting, since it appeared when adding an extra parameter, and thus was in contrast to our earlier example from \cref{fig:H0_w}, where a tension is relieved by introducing additional degrees of freedom. In this case, adding the curvature parameter drove the CMB and BAO contours apart, whereas enforcing a flat universe made them shift along their degeneracy lines to end up overlapping in the same spot (see \cref{fig:curvature_post_OmegaK_H0}).

Since then, there have been updates on both the CMB and BAO sides. Already with the 16th SDSS data release, the tension decreased by almost $\SI{0.5}{\sigma}$. But an even greater jump came with \Planck PR4 likelihoods. \CamSpec reduced the tension by over $\SI{1}{\sigma}$ and with \Hillipop the curvature tension drops further down to below the $\SI{1}{\sigma}$ threshold. Similarly, the curvature extension that previously was favored by \Plik on its own, performs no better than \LCDM for \CamSpec or is even slightly disfavored with \Hillipop. 

This shift towards flatness can also be seen in \cref{fig:curvature_post_OmegaK_H0}. Where previously the CMB and BAO contours did not even touch, there now is a clear overlap.\footnote{Note, however, that with the combination of \Planck~PR4~\cite{NPIPE}, ACT~DR6~\cite{ACT6_Louis2025} and SPT-3G~DR1~\cite{SPT3G_Camphuis2025} data, the curvature constraints shift back slightly, becoming more curved again compared to \Hillipop alone~\cite{Tristram2025_planck_act_spt}.}

\subsection{\texorpdfstring{\LCDM}{LCDM}}
\label{sec:lcdm}

With the release of the DESI data there has been an increased discussion of whether the CMB and BAOs are in tension with one another under \LCDM, and whether the evidence is pointing to needing a dynamical description of dark energy~\cite{DESI1_2024_cosmology,DESI2_2025_cosmology}.
Before diving deeper into the model comparison between \LCDM and \wwaCDM in \cref{sec:dark_energy}, we first review the tension just under \LCDM between CMB and BAOs on their own, and in combination with SNe.

\subsubsection{CMB versus BAO (versus lensing)}
\label{sec:cmb_bao}

\begin{figure}[tb]
    \centering
    \includegraphics[width=\columnwidth]{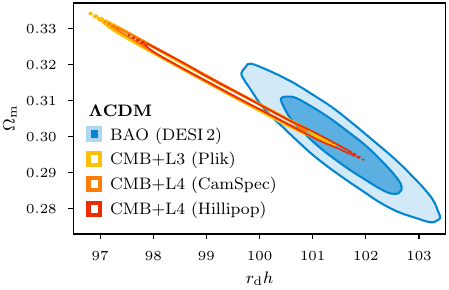}
    \caption{\label{fig:rdh_Om}
        Joint posterior of the matter density~\Omegam together with the product of the reduced Hubble parameter~$h$ and the scale of the sound horizon~$r_\mathrm{d}$ at the drag epoch. 
        For better visualization, and because of the high parameter degeneracy, only the \SI{95}{\percent} level is drawn for the CMB contours. Note how \CamSpec, in combination with lensing PR4, has the least overlap with DESI.
    }
\end{figure}

When comparing CMB to BAO data, it has become common to compare the joint posterior contours for the matter density~\Omegam, together with the product~$r_\mathrm{d}h$ of the reduced Hubble parameter~$h$ and the scale of the sound horizon~$r_\mathrm{d}$ at the drag epoch when baryons decouple from photons and are released from the radiation drag.
This parameter combination is convenient for BAO analyses, since BAO data effectively constrain two parameters (for more on the Bayesian model dimensionalities of the various data sets, see \cref{sec:dimensionalities,fig:dimensionalities}).
The $(r_\mathrm{d}h,\Omegam)$ posterior for the BAO data from DESI and for the three CMB likelihoods is shown in \cref{fig:rdh_Om}. The different axes of degeneracy for BAO and CMB data highlight how any tension cannot be reduced to a single parameter.

\begin{figure}[tb]
    \includegraphics[width=\columnwidth]{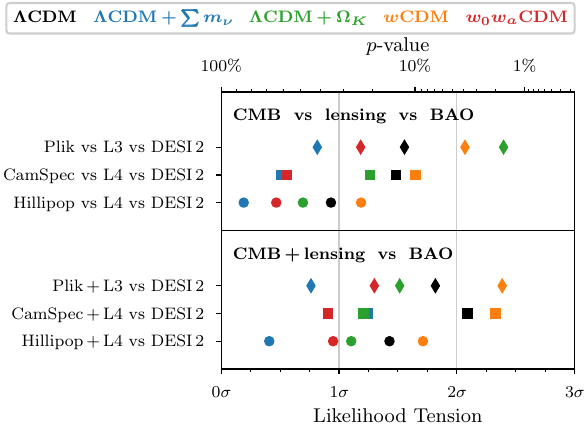}
    \caption{\label{fig:tension_cmb_lensing_bao}
        Comparison of the tension quantification between primary CMB, CMB lensing, and BAO data. The top block corresponds to a three-way comparison, whereas in the bottom block primary CMB and CMB lensing are compared jointly to the BAOs.
    }
\end{figure}

There are subtle differences in the tension statistics when comparing or combining more than two observables.
In \cref{fig:tension_cmb_lensing_bao} we compare the choice of a three-way tension quantification of primary CMB, CMB lensing, and BAO data to the tension that one gets from first combining primary CMB with CMB lensing and then computing the tension between this combination and BAOs.
Note that the latter can actually be inferred from the first through $\mathcal{R}_{A,B,C} = \mathcal{R}_{AB,C} \mathcal{R}_{A,B}$, together with the two-way tension from CMB versus lensing, which is why we did not make this type of distinction in \cref{fig:tension}.
However it is instructive to see how an initial combination of primary CMB and CMB lensing data leads to a slightly elevated tension with BAOs compared to the three-way approach, or compared to CMB versus BAO when CMB lensing is excluded, matching previous studies~\cite{DESI1_2024_cosmology,DESI2_2025_cosmology,GarciaQuintero2025_desi2_act6}.

For the most part, the tension statistics between CMB and lensing from \Planck versus BAOs from DESI are below $\SI{2}{\sigma}$. For \LCDM, the $\SI{2}{\sigma}$ threshold is only exceeded when using \CamSpec with lensing, where we find a slightly higher value of $\SI{2.1}{\sigma}$ (in agreement with previous findings~\cite{DESI2_2025_cosmology,Ye2025_desi_tension,Hamidreza2025_w0waCDM_desi_planck}). This can also be seen in \cref{fig:rdh_Om}, where \CamSpec combined with lensing PR4 shows the least overlap with the DESI contour.
The slight increase in tension when switching from \Plik to \CamSpec was recognized before~\cite{DESI2_2025_cosmology,GarciaQuintero2025_desi2_act6}. This comparison is sometimes implicitly conflated with a transition from \Planck~PR3 to PR4. However, our results indicate that this identification is incomplete: both \CamSpec and \Hillipop are based on \Planck~PR4, yet they yield measurably different tension statistics. In this sense, ``PR4'' does not uniquely fix the high\=/$\ell$ likelihood choice, and the residual likelihood dependence within PR4 is non-negligible for the inferred level of tension.

\begin{figure}[tb]
    \includegraphics[width=\columnwidth]{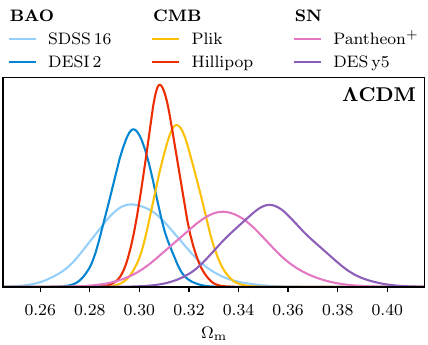}
    \caption{\label{fig:Omegam}
        Comparison of the matter density parameter under \LCDM as inferred from BAO, CMB, or SN data. Note the difference between the DESI~DR2 and DESy5 posteriors, which drives the tension when these two data sets are combined.
    }
\end{figure}

\subsubsection{CMB versus BAO versus SN}
\label{sec:cmb_bao_sn}

When adding supernovae into the mix, it should be noted that the tension under \LCDM between BAOs and SNe is actually stronger than between BAOs and the CMB. This is not surprising when looking at the main parameter driving the three-way (or four-way when viewing CMB lensing as distinct from CMB) comparison, the matter density parameter~\Omegam.
With a Bayesian model dimensionality of $\mathcal{d}\approx1$ (see also \cref{sec:dimensionalities,fig:dimensionalities}), the SN data are well summarized by \Omegam.
Its one-dimensional posteriors under \LCDM are shown in \cref{fig:Omegam}.
BAOs prefer a lower value of \Omegam, whereas SNe prefer a higher value, and the CMB falls somewhere in the middle.

\begin{figure}[tb]
    \includegraphics[width=\columnwidth]{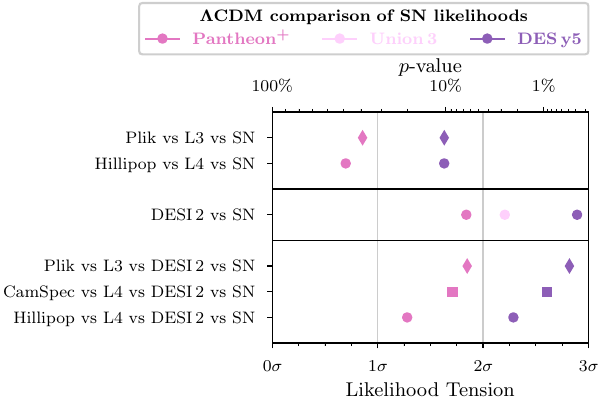}
    \caption{\label{fig:tension_desy5_pp}
        Selection of tension statistics from \cref{fig:tension}, highlighting the difference between the DESy5 and \PantheonPlus SN likelihoods under \LCDM.
    }
\end{figure}

Apparent from \cref{fig:Omegam} is the significant difference between the SN posteriors from \PantheonPlus and from DESy5. This difference contributes substantially to the tension statistics for all models (except for \wwaCDM, which is flexible enough to fit all data combinations, see \cref{sec:dark_energy}),
which is highlighted in \cref{fig:tension_desy5_pp}, a reduced version of \cref{fig:tension} focused on the DESy5--\PantheonPlus comparison.
There is a clear gap of about $\SI{1}{\sigma}$ throughout.

\begin{figure}[tb]
    \includegraphics[width=\columnwidth]{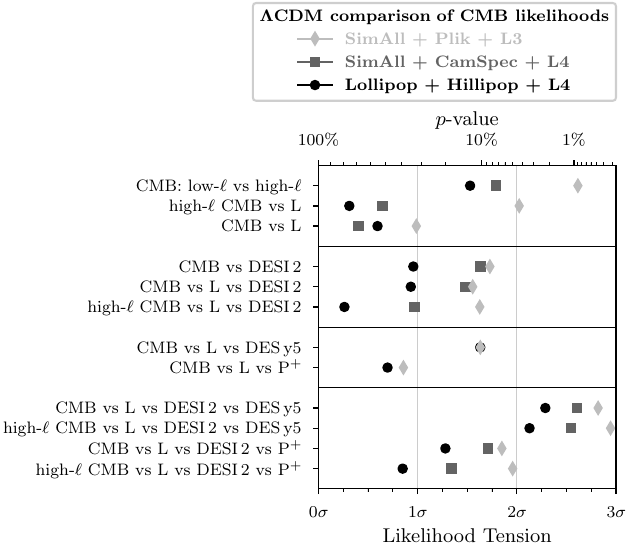}
    \caption{\label{fig:tension_plik_camspec_hillipop}
        Selection of tension statistics from \cref{fig:tension}, focused on the difference between the CMB likelihoods under \LCDM.
        There are relatively big jumps between PR3 (\Plik) and PR4 (\CamSpec and \Hillipop) for CMB internal consistency checks in the top block. There is also a non-negligible reduction in tension from \CamSpec to \Hillipop when combined with BAO data, despite both being based on the same \Planck release.
        For labels with ``high\=/$\ell$'', the corresponding low\=/$\ell$ likelihoods (\SimAll or \Lollipop, respectively) were dropped.
    }
\end{figure}

Not as dramatic as the differences between the two SN likelihoods, but still relevant, the systematic differences between the CMB likelihoods persist in the CMB, BAO, and SN combination.
As for SNe in \cref{fig:tension_desy5_pp}, we show a reduced version of the tension statistics for \LCDM in \cref{fig:tension_plik_camspec_hillipop}, with a focus on comparing the CMB likelihoods. There is a clear trend for a reduction in tension by going from \Plik to \CamSpec to \Hillipop. This is the strongest for the CMB internal comparison between low and high multipoles or between primary CMB and CMB lensing (as already mentioned in \cref{sec:cmb_internal}), with a notable jump when moving from the \Planck PR3 likelihood \Plik to the PR4 likelihoods \CamSpec or \Hillipop, suggesting that the reprocessing for PR4 did indeed clean up some systematics
(see also Ref.~\cite{Jense2025_planck_likelihood_choice} for a discussion of the parameter-level differences between PR3--\Plik and PR4--\CamSpec).
The difference between the \CamSpec and \Hillipop likelihoods is more subtle, as expected, considering that they are based on the same CMB maps.
However, in comparisons with data from DESI\,2, this difference still contributes about $\SI{0.5}{\sigma}$.
For comparisons with SN data, on the other hand, there is little difference in tension statistics.

These results reinforce a broader point already noted above: ``PR4'' does not uniquely specify the high\=/$\ell$ likelihood. Even holding the underlying maps fixed, moving between \CamSpec and \Hillipop induces shifts that are best interpreted as analysis-driven systematics associated with foreground treatment and nuisance modelling~\cite{CamSpec4_Rosenberg2022,Hillipop_Tristram2023}. We therefore view the \CamSpec--\Hillipop spread as a practical handle on the residual systematic uncertainty within PR4.

\subsection{Dark energy\texorpdfstring{\\}{ }(A cosmological constant tension?)}
\label{sec:dark_energy}

\subsubsection{\wCDM}

\begin{figure}[tb]
    \includegraphics[width=\columnwidth]{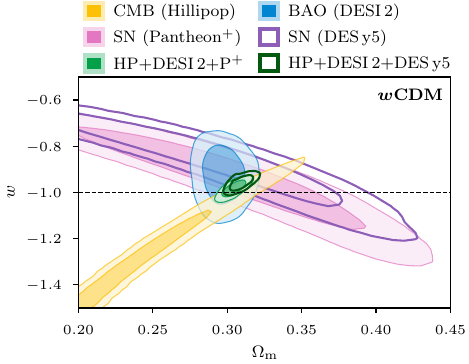}
    \caption{\label{fig:wCDM_post_cmb_vs_bao_vs_sn}
        Constraints on the \wCDM model for various data sets. 
        BAO and SN can resolve their \LCDM difference in \Omegam for $w>-1$; however, the CMB constraint pulls them back down.
        There is also a notable shift between the \PantheonPlus and DESy5 supernovae data.
    }
\end{figure}

A replacement of the cosmological constant~$\Lambda$ with a sampled equation-of-state parameter of dark energy~$w$ (constant in time), does not significantly reduce tensions or improve fits to overcome the Occam penalty from the enlarged parameter space, as seen in \cref{fig:tension_R,fig:tension,fig:model_comparison}. The only exception is when combining BAO with SN data, where tensions are minimal for a \wCDM cosmology. For DESy5 SN data, there is also a moderate preference for \wCDM over \LCDM in a direct model comparison.
The 2D contours in \cref{fig:wCDM_post_cmb_vs_bao_vs_sn} show this. The anti-correlation of the SN data between the equation-of-state parameter~$w$ and the matter density parameter~\Omegam enables the lower \Omegam value of BAOs to be achieved for a $w>-1$ value. 
However, as discussed in the previous section, when switching from DESy5 to \PantheonPlus the model preference for \wCDM is lost. Additionally, including CMB data pulls the joint contour back down towards $w=-1$, resulting in tension statistics similar to those under \LCDM, and gives a moderate to strong preference for \LCDM over \wCDM, such that an extension with just $w$ on its own is insufficient.

\subsubsection{\texorpdfstring{\wwaCDM}{w0waCDM}}

\begin{figure}[tb]
    \includegraphics[width=\columnwidth]{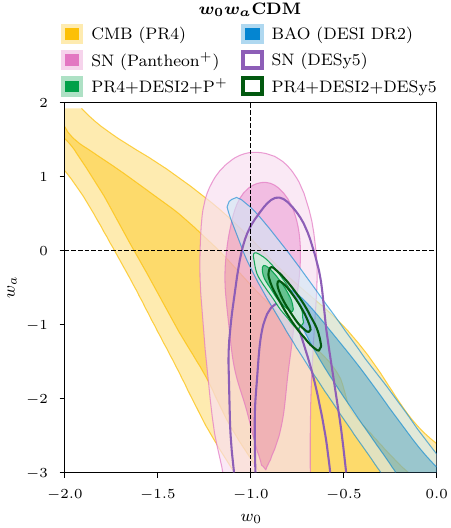}
    \caption{\label{fig:w0waCDM_post_ww}
        Constraints on the \wwaCDM model for various data sets. Note the significant shift between the \PantheonPlus and DESy5 supernovae data.
    }
\end{figure}

If we instead replace the cosmological constant~$\Lambda$ with a time-evolving description of dark energy $w(a)$---or equivalently $w(z)$ with respect to redshift~$z$---then the added flexibility of the model results in reduced tension statistics across a wide range of data combinations (see \cref{fig:tension_R,fig:tension}).
Particularly noticeable is the $\SI{1}{\sigma}$ to $\SI{1.5}{\sigma}$ gap between the \wwaCDM model and all the other extensions under a three-way comparison between CMB, BAOs, and SNe from DESy5.
This is also visible in the model comparison of \cref{fig:model_comparison}, where \wwaCDM is moderately preferred over \LCDM under DESI\,2\,+\,DESy5 with a log-evidence difference of ${\Delta\ln\mathcal{Z}\approx2.5}$, and strongly preferred under CMB\,+\,DESI\,2\,+\,DESy5 (independent of CMB likelihood) with a log-evidence difference of ${\Delta\ln\mathcal{Z}\approx4}$.
This is reflected in the joint CMB+BAO+SN contours in \cref{fig:w0waCDM_post_ww}, which for DESy5 are relatively far from the \LCDM point of $w_0=-1$ and $w_a=0$.

However, as already pointed out in \cref{sec:lcdm,fig:tension_desy5_pp}, this hinges on the use of the DESy5 SN data, and switching to the \PantheonPlus likelihood shifts the contours back towards \LCDM.
The significant gap in tension between \wwaCDM and the other models then shrinks considerably, and model comparisons completely switch from the moderate to strong preference for \wwaCDM to a moderate preference for \LCDM (log-evidence difference between $\Delta\ln\mathcal{Z}\approx-0.6$ for \CamSpec and $\Delta\ln\mathcal{Z}\approx-1.7$ for \Hillipop).

\subsubsection{DESy5 versus DES Dovekie}

During the writing of this paper Ref.~\cite{Popovic2025_Dovekie_calibration_uncertainties} released a reassessment of SN calibration uncertainties, leading to an update of the DES SN likelihood, referred to as ``DES Dovekie''~\cite{DESy5_2025_Dovekie_cosmo_update}. This update shifts the \Omegam posterior by over $\SI{1}{\sigma}$, to $\Omegam=0.330\pm0.015$, below the value from \PantheonPlus of $\Omegam=0.334\pm0.018$~\cite{DESy5_2025_Dovekie_cosmo_update,PantheonPlus_Brout2022}.

This confirms previous concerns about potential systematics in the DESy5 SN likelihood raised by Ref.~\cite{Efstathiou2025_supernovae}, and explains the qualitative reversal we observe when switching between SN likelihoods, as described in the previous sections.
We expect that any re-run with DES Dovekie would result in similar statistics to \PantheonPlus. Tensions for \LCDM would drop below the $\SI{2}{\sigma}$ threshold with comparable jumps for the other one-parameter extensions (neutrinos, curvature, or constant dark energy), closing the gap to the more flexible \wwaCDM model.
Similarly, in terms of model comparisons, the preference for \wwaCDM would switch to a slight preference for \LCDM, while the evidence of the one-parameter extensions remains mostly unaffected.

Since the strongest \wwaCDM-leaning shifts in our tension metrics and Bayesian evidence are driven by the DESy5 SN likelihood, the DES Dovekie update confirms that claims of a required update from \LCDM to \wwaCDM based on DESy5 were premature.


\subsection{Optical depth due to reionization\texorpdfstring{~$\tau_\mathrm{reio}$\\}{ }(A \texorpdfstring{$\tau$}{tau} tension?)}
\label{sec:tau}

\begin{figure}[tb]
    \includegraphics[width=\columnwidth]{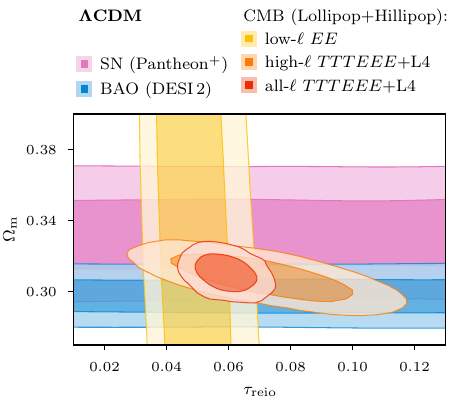}
    \caption{\label{fig:tau}
        Interplay between the optical depth due to reionization~$\tau_\mathrm{reio}$ and various data sets.
        While omission of the low\=/$\ell$ $EE$-polarization data indeed makes CMB and BAO data more compatible, this comes with a shift to lower \Omegam values, which drives it away from the \Omegam constraints from SN data.
    }
\end{figure}

\begin{figure*}
    \begin{minipage}[t]{\columnwidth}
        \includegraphics[scale=1.1]{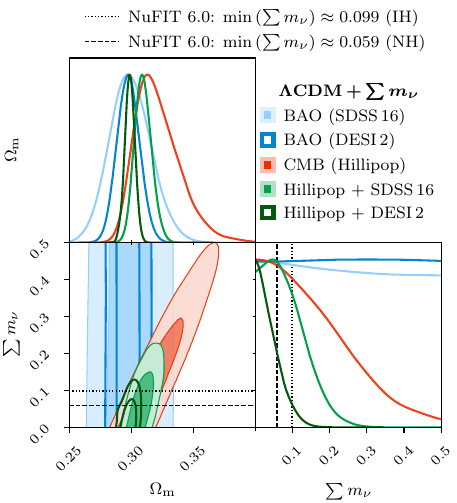}
        \caption{\label{fig:neutrino_cmb_vs_baos}
            Effect on the sum of neutrino masses~$\sum{m_\nu}$ of the BAO update from SDSS~DR16 to DESI~DR2.
            Combining \Planck~PR4 data with SDSS~DR16 results in a posterior peak for positive values of $\sum{m_\nu}$.
            This changes with DESI~DR2, where the posterior peaks at negative values, albeit less so than when combining with \Planck~PR3 or CamSpec~PR4 (see also \cref{fig:neutrino_cmbs_vs_bao_vs_sn}).
        }
    \end{minipage}
    \hfill
    \begin{minipage}[t]{\columnwidth}
        \includegraphics[scale=1.1]{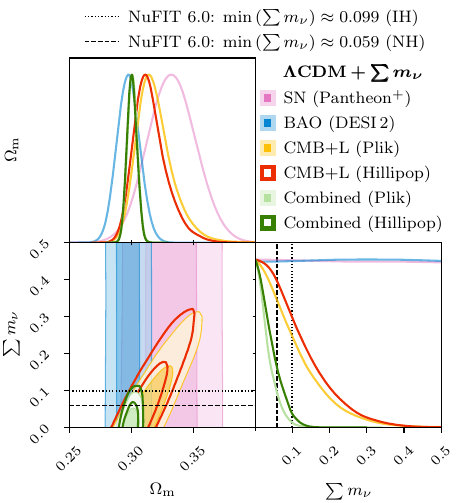}
        \caption{\label{fig:neutrino_cmbs_vs_bao_vs_sn}
            Effect on the sum of neutrino masses~$\sum{m_\nu}$ of the CMB update from \Planck~PR3 to \Planck~PR4.
            Note that \Planck~PR4 reduces the uncertainty for all other cosmological parameters compared to \Planck~PR3; only in the case of $\sum{m_\nu}$ does it have the effect of relaxing the constraint, which carries through to any combination with BAO and SN data.
        }
    \end{minipage}
\end{figure*}

Recent work has highlighted that mild CMB vs BAO inconsistencies can be rephrased in terms of the optical depth~$\tau_\mathrm{reio}$ due to reionization, in the sense that relaxing the low\=/$\ell$ $E$-mode polarization information that anchors $\tau_\mathrm{reio}$ can shift CMB+BAO inferences towards higher values \cite{Sailer2025_tau,Allali2025_tau}.

In our analysis, however, the baseline CMB--BAO tension is at most mild (and typically below $\SI{2}{\sigma}$, see \cref{fig:tension_cmb_lensing_bao}), with the inferred tension level depending noticeably on the specific CMB likelihood implementation.
In \cref{fig:tau} we nevertheless illustrate the same qualitative mechanism: omitting the low\=/$\ell$ $EE$ constraint from \Planck does indeed move the CMB posterior in the $(\tau_\mathrm{reio},\Omegam)$ plane in a direction that increases overlap with BAO data from DESI.
However, this shift is accompanied by a preference for lower \Omegam, which in turn worsens agreement with SN constraints.
We therefore do not interpret this as evidence for a standalone $\tau$-driven discrepancy in the full CMB+BAO+SN context, but rather as an indication of the sensitivity to how the large-scale polarization likelihood is treated.

This is also reflected in the tension statistics. 
In \cref{fig:tension_plik_camspec_hillipop} we show how the tension statistics change when we exclude the low\=/$\ell$ likelihoods and only use high\=/$\ell$ CMB data, both for the CMB--lensing--BAO and the CMB--lensing--BAO--SN combinations.
Without SN data, we can obtain a reduction of almost $\SI{0.7}{\sigma}$ from the exclusion of the low-$\ell$ likelihoods. Curiously, we only see this reduction for the PR4 likelihoods, not for \Plik, which starts out with the highest (albeit nonetheless less than $\SI{2}{\sigma}$) tension value.
However, with the SN data, any reduction diminishes considerably. Notably, the reduction is less for DESy5 (reduction of less than $\SI{0.2}{\sigma}$), where a tension would be more relevant (values over $\SI{2}{\sigma}$), than for \PantheonPlus, where any tension is negligible to begin with (values under $\SI{2}{\sigma}$).

Overall, we already do not see a necessity for a re-interpretation of a CMB and BAO tension in terms of a $\tau$ tension---a questionable premise that requires omitting the low\=/$\ell$ $E$-mode polarization information that directly anchors $\tau_\mathrm{reio}$---and, in light of SN data, such an interpretation does not appear helpful.

\subsection{Neutrino masses}

In our \LCDM baseline model we use the simplified approach to neutrinos of considering only a single massive neutrino with a mass of $m_\nu=\SI{0.06}{\eV}$. Making the assumption of a single massive neutrino can be viewed as an approximation of the normal neutrino hierarchy, with the mass close to the minimal sum one would obtain from the mass-squared differences of $\Delta{m}_{21}^2 = \SI{7.49(19)e-5}{\eV\squared}$ and $\Delta{m}_{31}^2=\SI{2.513(21:19)e-3}{\eV\squared}$, as reported by \NuFIT~\cite{NuFIT6}.

\Cref{fig:neutrino_cmb_vs_baos,fig:neutrino_cmbs_vs_bao_vs_sn} show the constraints we obtain from sampling the sum of three degenerate massive neutrinos, defining $M_\nu\equiv\sum{m_\nu}$, with an emphasis on the effect of a BAO update in \cref{fig:neutrino_cmb_vs_baos}, and an emphasis on the effect of the choice of CMB likelihood in \cref{fig:neutrino_cmbs_vs_bao_vs_sn}.


With SDSS~DR16 the posterior peaked at positive values of $M_\nu$, but the tighter constraints from DESI~DR2 push the peak to negative values, as already observed in Refs.~\cite{Green2025_negative_neutrino,Elbers2025_negative_neutrino}.
In Ref.~\cite{Elbers2025_negative_neutrino} there is a claim that adding SN data would push $M_\nu$ even lower than what is obtained from \Planck{}+DESI. However, we cannot confirm this. As clear from \cref{fig:neutrino_cmbs_vs_bao_vs_sn}, SN data pull the fits to higher values of \Omegam and hence to higher values of $M_\nu$ when combined with CMB data, going from the \SI{95}{\percent} upper bounds
\begin{align}
        \Plik + \mathrm{DESI\,2}: \sum{m_\nu} < \SI{0.069}{\eV}, \\
     \CamSpec + \mathrm{DESI\,2}: \sum{m_\nu} < \SI{0.067}{\eV}, \\
    \Hillipop + \mathrm{DESI\,2}: \sum{m_\nu} < \SI{0.081}{\eV}
\end{align}
for CMB and BAO data to
\begin{align}
        \Plik + \mathrm{DESI\,2} + \mathrm{P^+}: \sum{m_\nu} < \SI{0.073}{\eV}, \\
     \CamSpec + \mathrm{DESI\,2} + \mathrm{P^+}: \sum{m_\nu} < \SI{0.071}{\eV}, \\
    \Hillipop + \mathrm{DESI\,2} + \mathrm{P^+}: \sum{m_\nu} < \SI{0.089}{\eV}
\end{align}
for CMB, BAO, and SN data.
The upper bound on the sum of neutrino masses relaxes a bit with the newer CMB likelihoods.\footnote{Note that although \CamSpec shows the tightest bound, here, it was already shown that this bound relaxed from an earlier PR3 version of \CamSpec to the PR4 version used here (see Table 6 in Ref.~\cite{CamSpec4_Rosenberg2022}).} As pointed out in Ref.~\cite{Couchot2017_neutrinos_Alens}, this is directly related to the correlation between $M_\nu$ and the value of the lensing amplitude~$A_\mathrm{lens}$, which is much closer to its expected value of unity for the PR4 likelihoods~\cite{CamSpec4_Rosenberg2022,Hillipop_Tristram2023}.

In model comparisons, the neutrino extension performs similar or moderately weaker compared to \LCDM, indicating that any gain from a better fit is insufficient in offsetting the penalty from the slightly more complex model.

When looking at tension statistics on the other hand, the neutrino extension systematically reduces tension compared to \LCDM. This is most notable in the CMB+BAO combination with shifts of over $\SI{0.5}{\sigma}$, emphasising the push towards negative neutrino masses from Refs.~\cite{Green2025_negative_neutrino,Elbers2025_negative_neutrino}.

\section{Discussion and Conclusion}
\label{sec:conclusion}

``Concordance cosmology'' emerged as a slogan because the three independent data sets from CMB, BAOs, and SNe---each fairly degenerate in parameter space on its own---intersected in a common region of parameter space within uncertainties; close to flatness when allowing for curvature, and close to $w=-1$ when varying the dark-energy equation of state.
The aim of this work was to assess, in a unified Bayesian framework, how robust present-day claims of data-set inconsistency and preference for minimal extensions to \LCDM are when one varies the underlying data products (multiple \Planck data releases and likelihood implementations; BAO compilations spanning SDSS and DESI; and alternative SN likelihoods).
Several qualitative messages emerge, summarized in \cref{fig:tension_R,fig:tension,fig:model_comparison}.

First, updates from \Planck~PR3 to PR4 tend to improve internal CMB consistency, as quantified by low\=/$\ell$ versus high\=/$\ell$ likelihood splits and by primary-CMB versus CMB-lensing comparisons. While the details depend on the high\=/$\ell$ implementation, the overall trend is that PR4-era likelihoods reduce the level of internal disagreement relative to the most discrepant PR3 combinations, underscoring that some widely discussed ``tensions'' are sensitive to analysis choices and are not immutable properties of the CMB sky.

Second, we do not find compelling evidence for a standalone ``$\tau$ tension'' between low\=/$\ell$ polarization and BAOs, once the full CMB likelihood context is taken into account, despite recent discussion of the role of the reionization optical depth $\tau_\mathrm{reio}$ in CMB--BAO comparisons~\cite{Sailer2025_tau,Allali2025_tau}.

Third, varying $\sum m_\nu$ is not a preferred extension in our Bayesian model comparisons, though it can mildly reduce tensions for data combinations involving the CMB, for which \Omegam and $\sum{m_\nu}$ are positively correlated.

Fourth, the ``curvature tension'' is largely a PR3-era feature. Earlier analyses highlighted the combination of: (i)~a preference for closed geometries from certain CMB likelihood choices; and (ii)~an associated inconsistency with BAO and/or CMB lensing~\cite{DiValentino2020_curvature,Handley2021_curvature,DiValentino2021_curvature}. In our results, this pattern weakens substantially when moving to PR4 likelihoods, and it is further reduced when updating BAO inputs from SDSS~DR12 to DR16 or to DESI, consistent with the broader curvature-focused literature in the DESI era~\cite{Chen2025_cmb_bao_curvature}.

Fifth, claims of preference for evolving dark energy in combined CMB+BAO+SN analyses are not uniform across data products. In our analysis the strongest apparent preference for \wwaCDM arises for specific pairings, most notably \Planck~PR3 with DESy5, whereas switching to \Planck~PR4 and/or \PantheonPlus restores the \LCDM preference and reduces inferred tensions.
In this sense, our results support the following conservative conclusion: at present, a claim that \LCDM must be replaced by dynamical dark energy is premature when judged across reasonable choices of data products (as opposed to singling out particular data-set combinations).

There are multiple broader lessons that merit emphasis.
A useful interpretive point is that not all ``tension relief'' under model extensions is equally informative. There is a fundamental difference between the PR3-era curvature tension, on the one hand, and the cosmological-constant (or $H_0$) tension with preference for evolving dark energy, on the other hand.
The discussion around the curvature tension came up, because: (i)~CMB data alone preferred closed universes over flat universes even when penalizing for the additional parameter; and (ii)~the combination with BAO or lensing data was in tension with the curvature extension, but not with base \LCDM.
This was remarkable, because the addition of an extra parameter led to the tension, whereas naively we would expect additional parameters to make it easier to fit multiple potentially disparate data sets, i.e., we would expect an additional parameter generally to relieve tensions.
By contrast, many dark-energy extensions ease tensions primarily by degrading the constraining power of one data set. 
For example, when moving from \LCDM to \wCDM or \wwaCDM, the otherwise very predictive CMB constraints become substantially less informative, due to new degeneracy directions for parameters such as the Hubble parameter~$H_0$ or the matter density~\Omegam.
In such cases, the apparent reduction in tension actually reflects a loss of constraining power rather than a demonstrably improved simultaneous fit.

As emphasized in community discussions, experiments often optimize systematics control for a targeted parameter set within a baseline cosmology; when one combines data sets from different observables and enlarges the cosmological parameter space, the effective directions probed can differ substantially, and it is less clear a priori that the relevant systematics remain subdominant in the joint analysis. This consideration motivates continued end-to-end validation and cross-checks in the broader parameter spaces required for joint-probe inference.

Furthermore, it remains beneficial to maintain multiple independent analysis pipelines for flagship data sets. Comparisons across likelihood implementations and processing choices provide a practical handle on residual systematics and analysis-induced variance, especially when evidence differences are modest and when tensions sit near conventional ``interesting'' thresholds.
In the particular case of \Planck, the full-sky temperature anisotropies have reached the cosmic-variance limit on the largest angular scales, and will therefore remain a cornerstone CMB data set for the foreseeable future. 
Recent and forthcoming improvements are primarily complementary---higher-resolution ground-based measurements on small scales (ACT/SPT~\cite{ACT6_Louis2025,SPT3G_Camphuis2025}) and improved large-scale polarization (e.g.\ \LiteBIRD~\cite{LiteBIRD_PTEP}). 
For this reason, differences between \Planck likelihood implementations (\Plik, \CamSpec, \Hillipop), and potential future refinements (e.g.\ along the lines of \Cosmoglobe~\cite{BeyondPlanck_I,CosmoglobeDR1_I,CosmoglobeDR2_I}), are best viewed as a practical ``processing-induced uncertainty'' that should be propagated in joint-probe tension and model-comparison analyses.

Finally, experience across probes suggests that apparent discrepancies can evolve as analyses mature and systematics handling and calibration are revised. 
This was already shown in the weak-lensing literature with the gradual eroding of the clustering (or $\sigma_8$ or $S_8$) tension across successive analysis updates~\cite{Wright2025_clustering_tension}.
In this work, the same general pattern appears in several places: CMB updates from \Planck~PR3 to PR4, or BAO updates from SDSS~DR12 to DR16 improve consistency.
But even more striking, the DESy5 SN data responsible for the major tension in combination with CMB and BAOs---inciting claims for the end of the \LCDM standard model, and ultimately motivating our investigation---was recently updated to the DES Dovekie likelihood with revised calibration and systematics treatment.
This update brings the DES supernova constraints into close agreement with \PantheonPlus, which in our analysis shows neither a notable CMB+BAO+SN inconsistency nor a preference for \wwaCDM.
Taken together, these examples motivate caution against overly strong claims of new physics, especially as forthcoming BAO 
and SN 
data sets become increasingly precise and, for some parameters, approach the constraining power of the CMB.
In future comparisons of data combinations, the three statistical approaches illustrated in~\cref{fig:tension_R,fig:tension,fig:model_comparison} of this paper will be an important part of the analysis tool box.

\begin{figure*}[t]
    \centering
    \includegraphics[width=\textwidth]{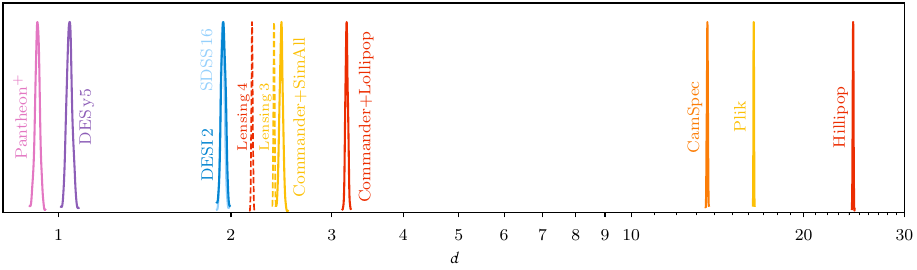}
    \caption{\label{fig:dimensionalities}
        Bayesian model dimensionalities~$\mathcal{d}$ of various data sets. There is approximately one constrained parameter in SN data and two in BAO data. The CMB lensing likelihoods and the low\=/$\ell$ primary CMB likelihoods constrain about two to three parameters. The high\=/$\ell$ CMB likelihoods come with a suite of nuisance parameters, such that they end up constraining more than just the six cosmological sampling parameters. Note that this only counts parameters \emph{constrained} by the data, which is why \Plik ends up with a lower dimensionality than \Hillipop despite sampling in principle more nuisance parameters (many \Plik nuisance parameters are prior dominated).
    }
\end{figure*}

\begin{acknowledgments}
This research was enabled in part by support provided by Calcul Qu\'ebec (\url{https://www.calculquebec.ca}) and the Digital Research Alliance of Canada (\url{https://www.alliancecan.ca}).
We gratefully acknowledge support from the CNRS/IN2P3 Computing Center (Lyon, France) for providing computing and data-processing resources needed for this work.
LTH is supported by a Postdoctoral Fellowship from the Centre National de la Recherche Scientifique~(CNRS) in France.
DS is supported by the Natural Sciences and Engineering Research Council of Canada.
\end{acknowledgments}

\Needspace*{9\baselineskip}
\section*{Data Availability}
\label{sec:data_availability}

The data that support the findings of this article (MCMC and nested sampling chains, post-processing and visualization code) are openly available on \textsc{Zenodo}~\cite{Zenodo2026_hergt_consistency}.
The \texttt{CosmoPower} emulators for each of our investigated cosmological models (\LCDM, \KLCDM, \nuLCDM, \wCDM, \wwaCDM) trained for and used in this article will be made available on \textsc{GitHub}.\footnote{\url{https://github.com/lukashergt/cosmopower4cobaya}}
We would appreciate citations to this article from any work that makes use of the released MCMC or nested-sampling chains, emulators, or derived products.

\appendix











\section{Dimensionalities}
\label{sec:dimensionalities}

To provide intuition for the effective constraining power of each likelihood, \cref{fig:dimensionalities} shows the Bayesian model dimensionality~$\mathcal{d}$ as defined in \cref{eq:model_dimensionality}.  Since $\mathcal{d}$ measures the posterior variance of $\ln\mathcal{L}$, it estimates the number of effectively constrained directions in parameter space (including nuisance directions where relevant).  This helps rationalize why SN and BAO are well summarized by roughly one and two parameters, respectively, while high\=/$\ell$ CMB likelihoods constrain additional nuisance combinations and therefore yield larger~$\mathcal{d}$.





\bibliographystyle{revtex/apsrev4-2-etal-titles}
\bibliography{references}

\end{document}